\documentclass[journal]{IEEEtran}
\usepackage{cite}
\usepackage{graphicx,amssymb,lineno}
\usepackage{amsmath,amsfonts,amssymb}

\usepackage{algorithm}
\usepackage{algorithmic}
\usepackage[usenames]{color}
\usepackage{float}
\usepackage{mathtools}
\usepackage{cite}
\usepackage{color}

\usepackage{graphicx,graphics,color,epsfig,subfigure,graphpap,rotate}
\usepackage{times, verbatim, subfigure, epsfig, graphicx, latexsym, amsmath}
\usepackage{url}
\usepackage{subfigure}

\begin{document}

\title{\huge{Content Distribution based on Joint V2I and V2V Scheduling in mmWave Vehicular Networks}}

\author{Lan~Su,
        Yong~Niu,~\IEEEmembership{Member,~IEEE},
        Zhu~Han,~\IEEEmembership{Fellow,~IEEE},
        Bo~Ai,~\IEEEmembership{Fellow,~IEEE},
        Ruisi~He,~\IEEEmembership{Senior Member,~IEEE},
        Yibing~Wang,
        Ning~Wang,~\IEEEmembership{Member,~IEEE},
        and Xiang~Su~\IEEEmembership{Member,~IEEE}

\thanks{Copyright (c) 2015 IEEE. Personal use of this material is permitted. However, permission to use this material for any other purposes must be obtained from the IEEE by sending a request to pubs-permissions@ieee.org.
This study was supported by National Key R\&D Program of China (2020YFB1806903): in part by the National Natural Science Foundation of China Grants 61801016, 61725101, 61961130391, and U1834210; in part by the National Key Research and Development Program under Grant 2021YFB2900301; in part by the State Key Laboratory of Rail Traffic Control and Safety (Contract No. RCS2021ZT009), Beijing Jiaotong University): and supported by the open research fund of National Mobile Communications Research Laboratory, Southeast University (No. 2021D09): in part by the Fundamental Research Funds for the Central Universities, China, under grant number 2020JBZD005: and supported by Frontiers Science Center for Smart High-speed Railway System; in part by the Fundamental Research Funds for the Central Universities 2020JBM089; in part by the Project of China Shenhua under Grant (GJNY-20-01-1). This work was partially supported by NSF CNS-2128368, CNS-2107216, Toyota and Amazon.}

\thanks{Lan Su is with the State Key Laboratory of Rail Traffic Control and Safety, Beijing Jiaotong University, Beijing 100044, China, and also with the Frontiers Science Center for Smart High-speed Railway System, Beijing Jiaotong University, Beijing 100044, China (e-mail: 19120118@bjtu.edu.cn).}

\thanks{Yong Niu is with the State Key Laboratory of Rail Traffic Control and Safety, Beijing Jiaotong University, Beijing 100044, China, and also with the National Mobile Communications Research Laboratory, Southeast University, Nanjing 211189, China (e-mail: niuy11@163.com).}

\thanks{Zhu Han is with the Department of Electrical and Computer Engineering, University of Houston, Houston, TX 77004 USA, and also with the Department of Computer Science and Engineering, Kyung Hee University, Seoul 446-701, South Korea (e-mail: zhan2@uh.edu).}

\thanks{Bo Ai, Ruisi He, and Yibing Wang are with the State Key Laboratory of Rail Traffic Control and Safety, Beijing Jiaotong University, Beijing 100044, China (e-mails: boai@bjtu.edu.cn; ruisi.he@bjtu.edu.cn; 18111034@bjtu.edu.cn).}

\thanks{Ning Wang is with the School of Information Engineering, Zhengzhou University, Zhengzhou 450001, China (e-mail: ienwang@zzu.edu.cn).}

\thanks{Xiang Su is with the Norwegian university of science and technology, Norway and University of Oulu, Finland (e-mail: xiang.su@ntnu.no).}
}

\maketitle

\begin{abstract}
With the explosive growth of vehicle applications, vehicular networks based on millimeter wave (mmWave) bands have attracted interests from both academia and industry. mmWave communications are able to utilize the huge available bandwidth to provide multiple Gbps transmission rates among vehicles. In this paper, we address the content distribution scheduling problem in mmWave vehicular networks. It has been challenging for all vehicles in the same network to complete content downloading due to the limited communication resources of roadside units (RSUs) and the high mobility of vehicles. We propose a joint vehicle-to-infrastructure (V2I) and vehicle-to-vehicle (V2V) scheduling scheme to minimize the total number of content distribution time slots from a global optimization perspective. In the V2I phase, the RSU serially transmits integrity content to vehicles, which are selected according to the vehicular network topology and transmission scheduling scheme. In the V2V phase, full-duplex communications and concurrent transmissions are exploited to achieve content sharing between vehicles and improve transmission efficiency. Performance evaluations demonstrate that our proposed scheme reduces the number of time slots and significantly improves system throughput when compared with other schemes, especially under large-size file transfers and a large number of vehicles.
\end{abstract}


\section{Introduction}\label{S1}

In recent years, the business potential of vehicle applications have increased with the development of Intelligent Transportation Systems (ITS), which requires new communication architectures with ultra-low latency and high throughput\cite{5G}. Therefore, millimeter wave (mmWave) communications with a data rate of up to 7 Gbps are a promising solution in vehicular networks\cite{Support}. mmWave vehicular networks have gained considerable attention from academia, industry, and standardization bodies\cite{Framework,Challenges,Challenges2}. Content distribution is crucial in ITS applications to assist vehicles in acquiring information \cite{Networking,Data,Taxis,survey}. Examples of content distribution may include, advertising companies periodically broadcast multimedia advertisements of local businesses in a city to vehicles driving the city on suburban highways (like a digital billboard), a traffic authority delivers real-time traffic information ahead, and disseminates an updated version of a local Global Positioning System (GPS) map \cite{CodeOn,Dynamic}. Such ITS applications can improve the safety factor of the transportation system, the utilization of road networks, and the convenience of travel.

In this paper, we address the crucial challenges of the content distribution in mmWave vehicular networks. Vehicle-to-infrastructure (V2I) and
vehicle-to-vehicle (V2V) communications are exploited for content distribution \cite{Overview}. V2I communication is an enabling technology for vehicles to access the Internet in real-time through roadside units (RSUs) to download content. However, the deployment of RSUs on the road is sparse with the constraints of geographical location and cost. The RSU is usually under heavy load due to the rapid growth of vehicles requesting and downloading content\cite{Chen}. Therefore, RSUs cannot provide data services for all requesting vehicles, which leads to content distribution failure. V2V communication leverages vehicle mobility and multi-hop relay among vehicles to expand the communication range. For vehicular networks, the high-speed mobility of the vehicular network results in a fast-varying channel environment. Hence, V2V communication has unpredictable transmission delays. It is challenging to meet the reliability and timeliness requirements of ITS applications\cite{D2D}. The two types of vehicular communications have some shortcomings, such as the intermittent connections of V2I communication and the unreliability of V2V communication \cite{Road}. In general, content distribution relying solely on one of the above-mentioned communication methods can only provide users with lightweight services, such as downloading video clips and browsing web pages. It does not allow for the downloading of large-size files in a sufficient time \cite{hoc}. Therefore, we propose a content distribution scheme based on joint V2I and V2V scheduling to reduce the burden of the RSU, download large-size files, and extend the time for vehicles to receive the content.

To improve system throughput of content distribution in mmWave vehicular networks, we consider joint V2I and V2V communications to study content distribution scheduling. The content distribution can be divided into two phases, when a large number of vehicles send the same content download request to the RSUs. In the V2I phase, the RSU serially transmits the integrity content to the vehicles, which are selected according to the vehicular network topology. In the V2V phase, full-duplex (FD) communications and concurrent transmissions are exploited to improve transmission efficiency. The contributions of this paper can be summarized as follows:

\begin{itemize}
\item We propose a joint V2I and V2V scheduling scheme to minimize the total number of content distribution time slots from a global optimization perspective. The resource allocation of the RSU and the content forwarding of the V2V phase are two linked processes that jointly affect the number of time slots for content distribution.

\item We propose a joint V2I and V2V scheduling algorithm for the highly dynamic vehicular network topology in one-way multi-lane scenario. In the V2I phase, we propose a transmission path selection algorithm based on utility function to select the vehicles communicating with RSU. The utility function is designed based on minimizing the number of time slots for vehicles to complete content downloading. It fully considers the vehicular network topology and the transmission scheduling between vehicles. Further, in the V2V phase, we propose a scheduling algorithm based on FD communications and concurrent transmissions to improve transmission efficiency.

\item We conduct extensive performance evaluations of the proposed scheme under various system parameters. Results demonstrate that compared with the First come First serve (FCFS) cooperation scheme, the proposed scheme reduces the transmission time by about 40.6$\%$ and increases the system throughput by 59.2$\%$.
\end{itemize}

The rest of this paper is organized as follows. Section~\ref{S2} introduces the related work. In Section~\ref{S3}, we provide the system model, including the V2I channel model, V2V channel model, and antenna pattern. In Section~\ref{S4}, we formulate the content distribution problem with the objective of minimizing the number of transmission time slots. In Section~\ref{S5}, the transmission path selection algorithm in the V2I phase is presented, as well as the full-duplex concurrent transmission scheduling algorithm in the V2V phase. In Section~\ref{S6}, extensive simulations are performed to evaluate the proposed scheme. Finally, we conclude the paper in Section~\ref{S7}.

\section{Related Work}\label{S2}

There is much relevant literature on content distribution in vehicular networks. Some works focused on the V2I phase of cooperative content distribution without exploiting V2V communications \cite{WiFi-Based,MMCD}. In \cite{WiFi-Based}, an access points' collaborative content distribution system is proposed to improve the diversity of information circulating in vehicular networks. To reduce mean delivery delay, Dong \emph{et al.} \cite{MMCD} proposed max-throughput and min-delay cooperative downloading, using a cooperative downloading algorithm in vehicular networks. Some works focused on the V2V phase of cooperative content distribution in vehicular networks \cite{Dynamic,Multi-Hop,Wang}. To minimize the overall transmission delay, Lv \emph{et al.} \cite{Multi-Hop} proposed a blockage avoidance based multi-hop sensor data dissemination scheme in a centralized manner. In \cite{Dynamic}, a cooperative half duplex scheme based on alliance formation game is proposed to minimize the transmission delay of popular content distribution. To improve the transmission efficiency, Wang \emph{et al.} \cite{Wang} proposed a cooperation FD scheme based on coalition formation game to address the popular content distribution of mmWave vehicular networks. However, the proposed schemes in \cite{Dynamic} and \cite{Wang} are established under ideal transmission in the V2I phase, which is unrealistic in practice. This is because the performance decreases when the content size increases.

Only studying the content distribution scheduling of V2I or V2V communications lead to unsatisfactory transmission performance. High transmission delay and low transmission success rate cannot meet the demand of vehicles for content downloading. To improve the system throughput and reduce transmission time, Chen \emph{et al.} \cite{Chen} proposed a joint resource allocation and scheduling approximate algorithm by combining the resource allocation in the access point's coverage with the transmission scheduling in the Internet link-hole area. To improve fairness among vehicle users, Zheng \emph{et al.} \cite{Graph} proposed a bipartite-graph-based scheduling scheme to allocate the V2I and V2V links for both single-hop and dual-hop communications. However, the radio resource is allocated equally to each link, which is not optimal for the performance of content distribution. Thus, a two-dimensional-multi-choice knapsack problem-based scheduling scheme is proposed to select the coordinator vehicles for the sinking vehicle in \cite{Joint}. In \cite{Joint}, the authors focus on allocating radio resources to V2V and V2I links. However, the proposed schemes in \cite{Chen,Graph,Joint} analyze the content distribution of V2I and V2V cooperative communications in a low-frequency network, not in the mmWave band. There are generally a few existing studies on solving the problem of content distribution in vehicular networks by combining mmWave and joint V2I and V2V communications. Besides, they do not consider how to minimize the number of time slots with all the downloading requests of vehicles being completed, resulting in some vehicles unable to complete content downloading. Hence, to ensure that each vehicle completes content downloading, we propose a content distribution scheme based on joint V2I and V2V scheduling when multiple vehicles request the same content.

\section{System Model}\label{S3}

\begin{figure}[t]
\begin{center}
\includegraphics*[width=1.0\columnwidth,height=1.6in]{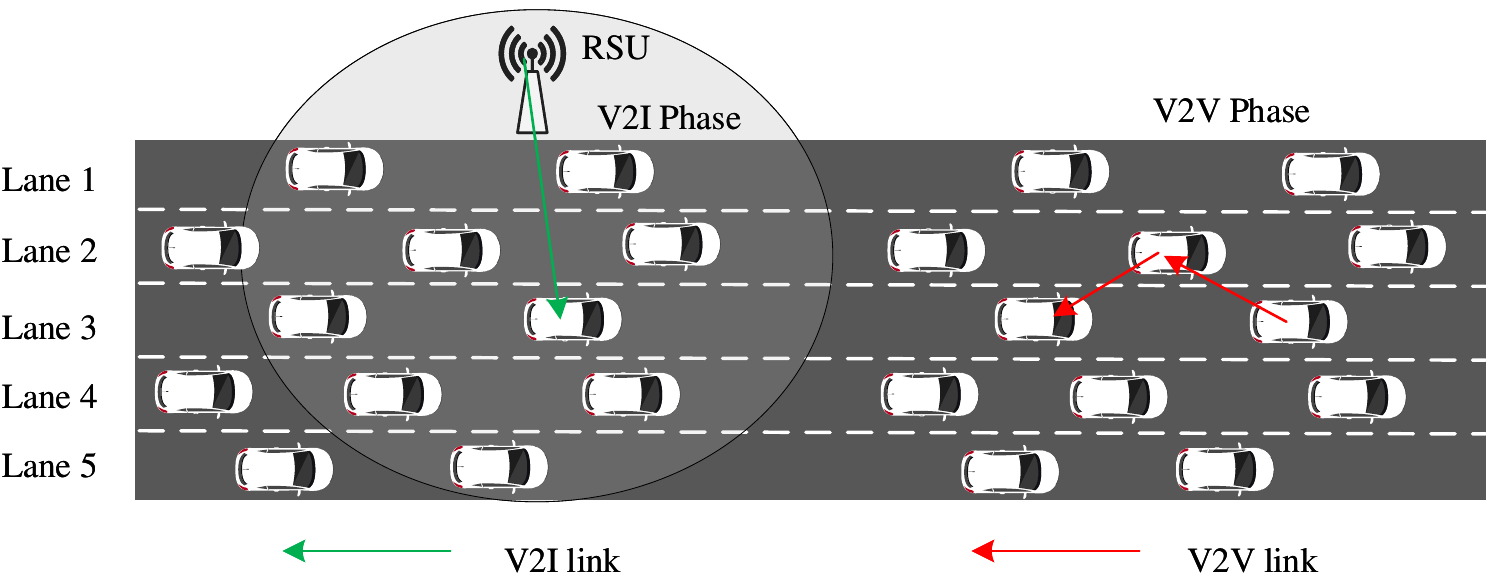}
\end{center}
\caption{System model of content distribution in mmWave vehicular networks.}
\label{fig1}
\end{figure}

The system model of content distribution in mmWave vehicular networks is shown in Fig. \ref{fig1}, where an RSU is deployed on the roadside and $N$ vehicles are on a one-way five-lane road. All vehicles require the same content when they pass through the RSU's coverage with the same constant speed. Let $\mathcal{V} = \{ {V_1},{V_2}, \ldots ,{V_N}\}$ denote the set of vehicles that require the same content, $\left| \mathcal{V} \right| = N$. The content distribution process can be divided into two phases: the V2I phase and V2V phase \cite{V2X1,CVCG}. In the V2I phase, only some vehicles receive a large-size file because of the RSU's limited coverage and vehicles' high speed. Furthermore, in the V2V phase, vehicles download the content by sharing data among the different vehicles.

To support generality, we assume any content requested by vehicles are available at the RSU, and users can obtain the contents in advance via the Internet by data prefetching methods \cite{Offloading}. We assume that the RSU can deal with the requests only one at a time, and the content size does not exceed the amount of data that the vehicle can download from the current RSU \cite{MMCD}. All vehicles can transmit and receive signals simultaneously, but each link carries out up to two-hop communications in the V2V phase. Moreover, we assume that vehicle trajectory is known, and all vehicles can obtain accurate information including positions, velocities, directions, etc. from GPS sensors \cite{Scenarios}.

To clarity the illustration, the transmission time is divided into $n$ equal time slots, and the total number of time slots with all the downloading requests of vehicles being completed is not greater than $n$. Let $\mathcal{T} = \{1,2, \ldots ,n\}$ denote the set of time slots, $\left| \mathcal{T} \right| = n$. In the remaining part of this paper, the number of time slots is used to measure the duration of the transmission time.

The V2I channel model, the V2V channel model, and the antenna model of the proposed scheme are presented in the following subsections in detail.

\subsection{V2I Channel Model}\label{S3-1}

For the V2I channel, the distance between the RSU and each vehicle is constantly changing because of the high-speed mobility of vehicles. For simplicity, we assume that the channel is unchanged in a time duration of one slot $\Delta$.
The transmitting signal from the RSU can only be received by the vehicles within the RSU's coverage. Since the RSU mmWave antenna is installed at a high position (compared with the height of vehicles mmWave antenna), there is scarcely an obstacle between the RSU and vehicles. Therefore, we assume that there is only a line-of-sight (LOS) link between them \cite{Directional}. In this way, the received signal-to-noise ratio (SNR) for vehicle $i$ within the RSU's coverage is given as
\begin{equation}
SN{R_i} = \left\{ {\begin{array}{*{20}{c}}
{\frac{{k{P_t}{G_t}{G_r}d_i^{ - \tau}}}{{{N_0}W}},\hspace{0.4cm} {d_i} \le {R_r},}\\
{0,\hspace{2cm}{d_i} > {R_r},}
\end{array}} \right.
\label{eq1}
\end{equation}
where ${P_t}$ is the RSU's transmission power, $k$ is a constant coefficient and proportional to ${(\frac{\lambda}{4\pi})^2}$ ($\lambda$ denotes the wave length), ${G_t}$ is the transmit antenna gain of RSU, ${G_r}$ is the receive antenna gain of vehicle $i$, $\tau$ is the path loss exponent, ${N_0}$ is the one-sided power spectral density of white Gaussian noise, $W$ is the channel bandwidth, ${d_i}$ is the distance between the RSU and vehicle $i$, and ${R_r}$ is the coverage of the RSU.

The transmission rate between the RSU and the driving vehicle is highly dynamic, because the distance between the RSU and the driving vehicle is changing. The maximum achievable transmission rate of vehicle $i$ can be given according to the Shannons channel capacity by ${R_i} = W{\log _2}\left( {1 + SN{R_i}} \right)$. For simplicity, we assume that the average transmission rate of each time slot is equal to the instantaneous rate. The detailed implemental process of the rate modeling can refer to \cite{Cooperative}. The average transmission rate in each time slot is computed by the integral of the link rate over time and is given as
\begin{equation}
\begin{array}{l}
R_i^t = \frac{1}{\Delta }\int_{t - 1}^t {{R_i}\left( t \right)} dt\\
\hspace{0.5cm} \buildrel \Delta \over = \frac{1}{\Delta }\int_{{\varphi _{t - 1}}}^{{\varphi _t}} {{R_i}\left( {{d_i}(t)} \right)} \frac{{{d_i}(t)d\varphi }}{{v\cos \varphi }}\\
\hspace{0.5cm} = \frac{1}{\Delta }\int_{{\varphi _{t - 1}}}^{{\varphi _t}} {{R_i}\left( {\frac{{{d_{lr}}}}{{\cos \varphi }}} \right)} \frac{{{d_{lr}}d\varphi }}{{v{{\cos }^2}\varphi }}\\
\hspace{0.5cm} = \frac{1}{\Delta }\int_{{\varphi _{t - 1}}}^{{\varphi _t}} {{{\log }_2}\left[ {1 + \frac{{k{P_t}{G_t}{G_r}}}{{{N_0}W}}{{\left( {\frac{{{d_{lr}}}}{{\cos \varphi }}} \right)}^{ - n}}} \right]} \frac{{W{d_{lr}}}}{{v{{\cos }^2}\varphi }}d\varphi
\end{array}
\label{eq2}
\end{equation}
where $t$  is the time slot, $\Delta$ is the time duration of one slot, $v$ is the speed of a vehicle, $\varphi$ is the complementary angle of the angle between the received signal direction of the vehicle and its driving direction, and ${d_{lr}}$ is the vertical distance between the RSU and vehicle $i$ when vehicle $i$ is driving on lane $l$.

To guarantee transmission links quality, quality of service (QoS) requirements of V2I links are given, i.e., the actual signal to interference plus noise ratio (SINR) of the link must be larger than the given SINR threshold. $th_{min}$ is defined as the SINR threshold of all links. However, since the RSU serially transmits the integrity content to the vehicles, there is no interfering signal posed by any vehicle. Therefore, the condition that the V2I link from the RSU to vehicle $i$ can be successfully transmitted can be expressed as $SN{R_i} \ge t{h_{min}}$.

\subsection{V2V Channel Model}\label{S3-2}

For the V2V channel, we assume that all vehicles have the same transmission power $P_v$ and communication range $R$, and $R < {R_r}$. The RSUs are assumed to cover wider ranges, because they have more transmission power than vehicles \cite{Throughput}. We assume that all vehicles drive at a constant speed \cite{MMCD,V2V/V2I}. So the distance and transmission rate between vehicles remain unchanged. We denote the directional link from vehicles $i$ to $j$ by $(i,j)$. Besides, we assume the V2V links exist only between neighboring vehicles with LOS links, the received power at vehicle $j$ for $(i,j)$ is given as
\begin{equation}
P{r_{i,j}} = \left\{ {\begin{array}{*{20}{c}}
{k{P_v}{G_t}(i,j){G_r}(i,j)d_{i,j}^{ - \tau},\hspace{0.4cm}{d_{i,j}} \le R,}\\
{0,\hspace{3.7cm}{d_{i,j}} > R,}
\end{array}} \right.
\label{eq3}
\end{equation}
where ${P_v}$ is the vehicle's transmission power, ${G_t}(i,j)$ is the transmit antenna gain of the link between vehicles $i$ and $j$, ${G_r}(i,j)$ is the receive antenna gain of the link between vehicles $i$ and $j$, and $d_{i,j}$ is the distance between vehicles $i$ and $j$.

Due to FD communications, self interference (SI) needs to be considered in the V2V phase. For the SI problem, many cancellation schemes have been proposed so far, but the existing schemes cannot completely eliminate SI. We assume that appropriate SI cancellation technology is employed, and then the amount of remaining self interference (RSI) can be represented by the transmission power of the link receiver in the calculation \cite{Full1,Full2}. If vehicle $j$ is the receiver of link $(i, j)$ and the transmitter of link $(j, f)$ at the same time, the RSI must exist at vehicle $j$ and can be denoted by ${\beta _j}{P_{{v_{j,f}}}}$ $(f\in\mathcal{V})$, where $\beta_j$ represents the SI cancelation level at vehicle $j$ and is a non-negative parameter. For each vehicle $i\in\mathcal{V}$, we define a binary variable $b_i^t$ to indicate whether vehicle $i$ is both a transmitter and a receiver in time slot $t$. If that is the case, we have $b_i^t = 1$; otherwise, $b_i^t = 0$.

Due to the limited communication range of vehicles, the interference between links cannot be neglected when concurrent transmissions are exploited in the V2V phase. For links $(u, f)$ and $(i, j)$, the received interference power at vehicle $j$ from vehicle $u$ can be calculated as
\begin{equation}
{I_{ufij}} = \left\{ {\begin{array}{*{20}{c}}
{\rho k{P_v}{G_t}(u,j){G_r}(u,j)d_{u,j}^{ - \tau},\hspace{0.3cm}{d_{u,j}} \le R,}\\
{0,\hspace{4.0cm}{d_{u,j}} > R,}
\end{array}} \right.
\label{eq4}
\end{equation}
where $\rho$ denotes the multi-user interference (MUI) factor related to the cross correlation of signals from different links \cite{UWB,STDMA}, and $d_{u,j}$ denotes the distance between vehicles $u$ and $j$.

The set of links transmitted simultaneously in time slot $t$ is denoted by $\mathbb{C}_{t}$. The interference power of the link from vehicles $i$ to $j$ can be calculated as
\begin{equation}
{I_{i,j}} = \sum\limits_{(u,f) \in {\mathbb{C}_t}\backslash \{ i,j\}} {{I_{ufij}}}.
\label{eq5}
\end{equation}
Thus, the SINR at the receiver of link $(i,j)$ in time slot $t$ can be calculated as
\begin{equation}
\begin{array}{l}
{\Gamma _{i,j}}(t) = \frac{{k{P_v}{G_t}(i,j){G_r}(i,j)d_{i,j}^{ - \tau}}}{{{N_0}W + \rho \sum\limits_{(u,f) \in {\mathbb{C}_t}\backslash \{ i,j\} } {k{P_v}{G_t}(u,j){G_r}(u,j)d_{u,j}^{ - n} + b_j^t{\beta _j}{P_v}} }}.
\end{array}
\label{eq6}
\end{equation}

In the V2V phase, the achievable data rate of link $(i,j)$ in time slot $t$ can be estimated according to the Shannons channel capacity as $R_{i,j}^t = W{\log _2}\left( {1 + {\Gamma _{i,j}}(t)} \right)$. The condition that the V2V link from vehicles $i$ to $j$ can be successfully transmitted can be expressed as ${\Gamma _{i,j}}(t) \ge t{h_{min}}$.

\subsection{Antenna Pattern}\label{S3-3}

In mmWave V2I and V2V communications, for tractability of analyzing the directivity gains of the antennas after beam alignment, the mmWave directional antenna pattern of RSUs and vehicles can be approximated by a two-dimensional ideal sectorial antenna model. Therefore, we adopt the ideal sectorial antenna model proposed in \cite{Beamwidth}, and the correspondence between the wave angle and the antenna gain can be given as
\begin{equation}
G(\theta) = \left\{ {\begin{array}{*{20}{c}}
{\frac{{2\pi  - (2\pi  - \phi )g}}{\phi },\hspace{0.8cm}\left| {{\theta}} \right| \le \frac{\phi }{2},}\\
{g,\hspace{2.4cm} {\rm otherwise},}
\end{array}} \right.
\label{eq7}
\end{equation}
where $\theta$ is the alignment error angle between vehicles or between the RSU and vehicles, $\phi$ is the angle of the half-power beamwidth, and $g$ is the sidelobe power, which is between at 0 and 1. The antenna pattern is used to calculate the transmitter antenna gains and receiver antenna gains in Equations (1)--(6).

\section{PROBLEM FORMULATION}\label{S4}

\begin{figure}[t]
\begin{minipage}[t]{1\linewidth}
\centering
\includegraphics[width=8cm]{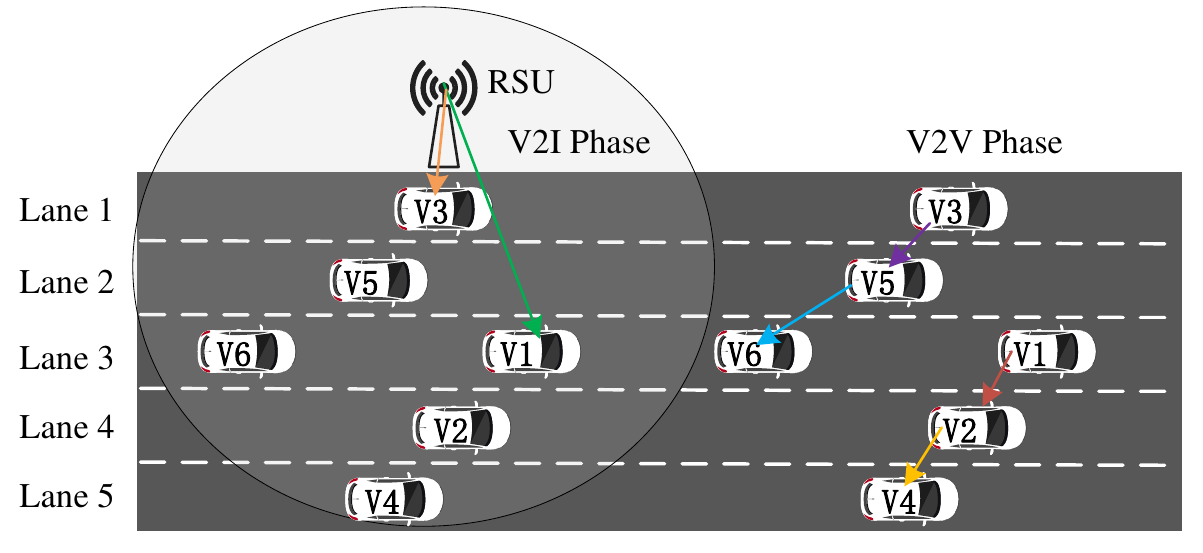}
\centerline{\small (a) The mmWave vehicular network topology}
\end{minipage}%
\\
\begin{minipage}[t]{1\linewidth}
\centering
\includegraphics[width=8cm]{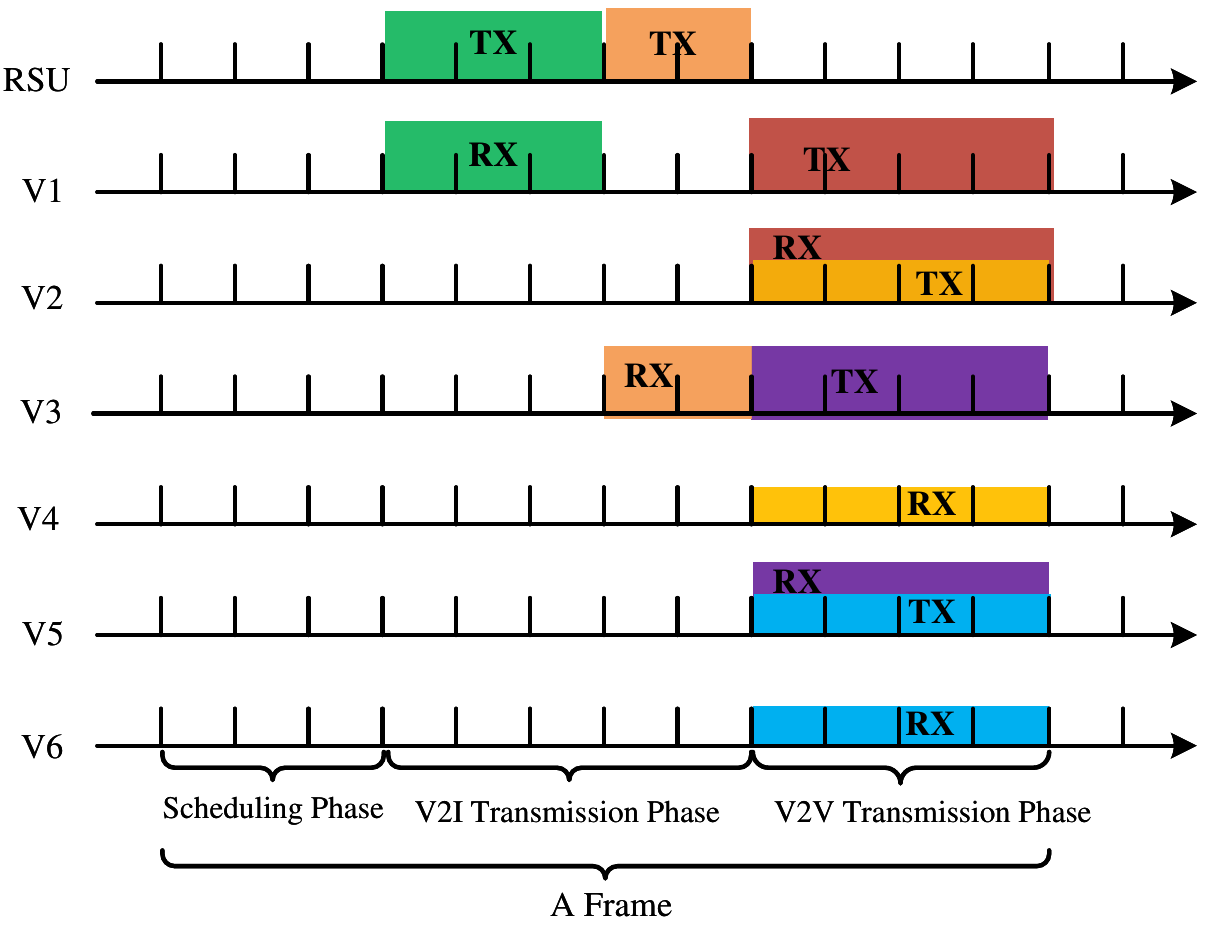}
\centerline{\small (b) Time-line illustration of the proposed scheme}
\end{minipage}
\begin{minipage}[t]{1\linewidth}
\centering
\includegraphics[width=8cm]{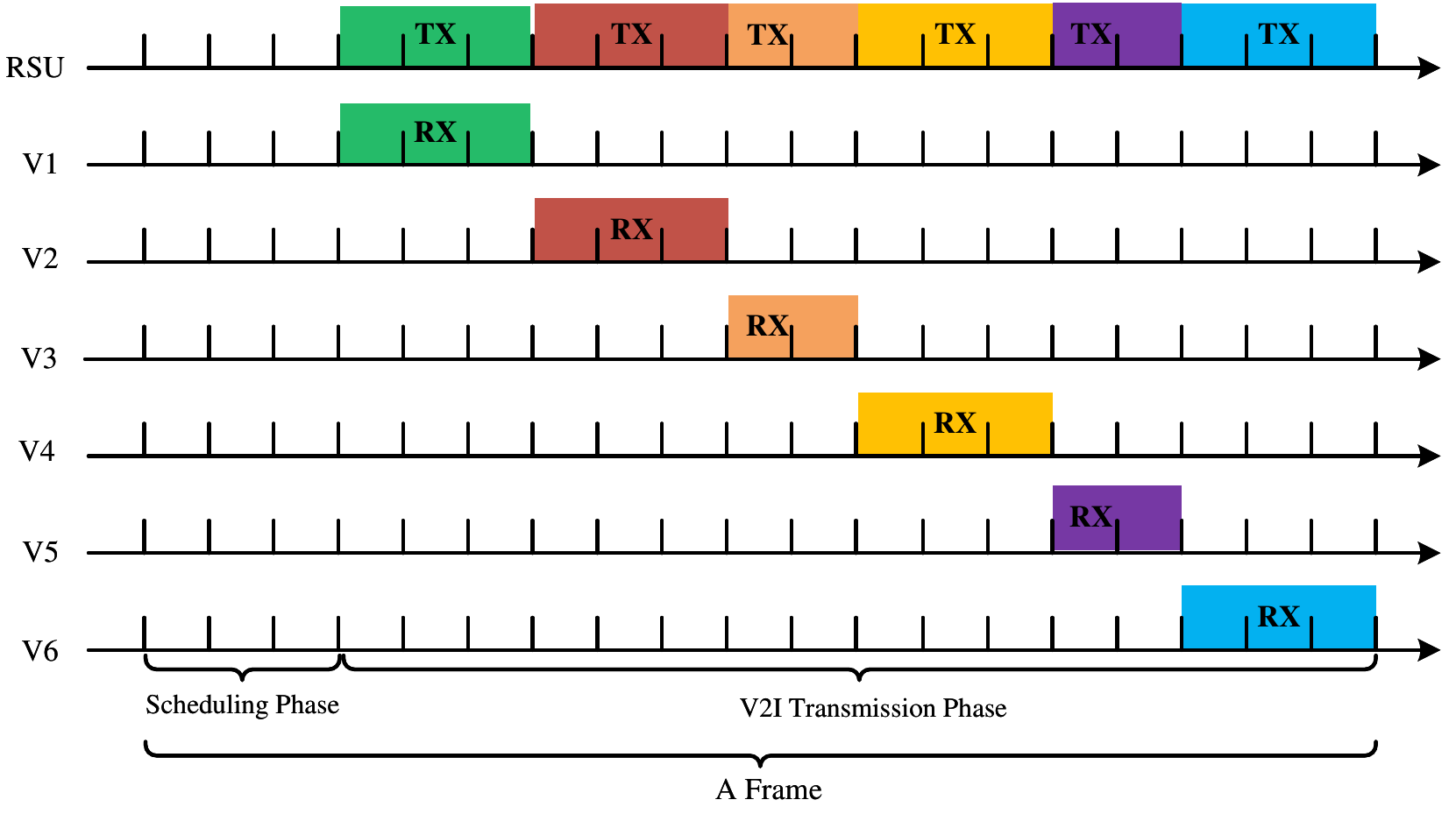}
\centerline{\small (c) Time-line illustration of the TDMA scheme}
\end{minipage}%
\\
\caption{An example of content distribution operation in a mmWave vehicular network of an RSU and multiple vehicles.}
\label{fig2} 
\vspace*{-3mm}
\end{figure}

When multiple vehicles request the same content, we consider joint V2I and V2V communications and take them as a whole to study the content distribution schedule. This work aims to minimize the number of time slots for all vehicles to successfully complete content downloading. Moreover, the main challenge is, from a time slot number minimization perspective, how to select the vehicles to download the content in the V2I phase to effectively be exploited by the content forwarding in the V2V phase.

In content distribution scheduling, time is divided into a sequence of non-overlapping frames \cite{Ding}. There are two phases in each frame, including the scheduling phase and transmission phase. In the scheduling phase, the RSU determines the content distribution scheduling scheme according to the V2I transmission path selection algorithm and the V2V FD concurrent transmission scheduling algorithm. In the transmission phase, the RSU serially transmits the integrity content to vehicles according to the scheduling scheme. Further, FD communications and concurrent transmissions are exploited to improve transmission efficiency in the V2V phase. To minimize the number of transmission time slots, it is necessary to consider the distance between the RSU and vehicles and the effectiveness of the content forwarding in the V2V phase, when the RSU selects a vehicle to download content from.

As shown in Fig. \ref{fig2} (a), we consider a scenario that six vehicles requiring the same content pass through the RSU's coverage at the same constant speed on a one-way five-lane road, and the serial number of the vehicles are arranged according to the sequence of vehicles entering the RSU's coverage. As shown in Fig. \ref{fig2} (b), in the V2I phase, the RSU selects vehicles $V_1$ and $V_3$ to download the content according to the location of vehicles. The RSU transmits to vehicle $V_1$ for three time slots, and the RSU transmits to vehicle $V_3$ for two time slots. Therefore, the content distribution uses five time slots in the V2I phase. In the V2V phase, there is a pairing in the transmission phase, and multiple links are activated simultaneously for concurrent transmissions. We obtain the transmission paths as path ${V_1} \to {V_2} \to {V_4}$ and path ${V_3} \to {V_5} \to {V_6}$. The paper focuses on using the FD communications mode, rather than designing the FD media access control \cite{MAC,RCTC,Nam}. For simplicity, in the three-node FD transmission ${V_1} \to {V_2} \to {V_4}$, we assume that when $V_2$ receives data from $V_1$, it can transmit the data to $V_4$ in parallel \cite{Duplex1}.
In the first two-hop FD transmission link, vehicle $V_1$ transmits to vehicle $V_2$ for four time slots, and vehicle $V_2$ transmits to vehicle $V_4$ concurrently for four time slots. In the second two-hop FD transmission link, vehicle $V_3$ transmits to vehicle $V_5$ for four time slots, and vehicle $V_5$ transmits to vehicle $V_6$ concurrently for four time slots. Therefore, the content distribution uses four time slots in the V2V phase. To complete the content distribution for six vehicles, the schedule needs nine time slots. As shown in Fig. \ref{fig2} (c), if we adopt the serial time division multiple access (TDMA) scheme, where the RSU transmits to six vehicles serially, we will need at least sixteen time slots to complete the content distribution. Therefore, in the same network topology, we can observe that the different resource allocation of the RSU will affect the multi-source forwarding of content in the V2V phase, and the content forwarding in the V2V phase largely determines the effectiveness of resource allocation of the RSU.

To sum up, the resource allocation of the RSU and the content forwarding of the V2V phase are two linked processes, which jointly affect the number of time slots for content distribution. We assume that when the first vehicle enters the coverage of the RSU, $t = 1$, and the set of vehicles that complete the content downloading in the V2I phase is denoted by $\mathcal{V}_{V2I}$. Each vehicle requires the same content, and the content size is denoted by $D$. The time for a vehicle to pass through the RSU's coverage on $l$ lane is ${t^l} = \frac{{2\sqrt {R_r^2 - d_{lr}^2} }}{v}$, $l = \{ 1,2,3,4,5\}$ and the content downloading can be completed within the time $t^l$. We define a binary variable $c_i^t$ to indicate whether the RSU communicates with the vehicle $i$ in time slot $t$. If that is the case, we have $c_i^t = 1$; otherwise, $c_i^t = 0$. Hence, the number of time slots for content distribution in the V2I phase is ${t_{V2I}} = \sum\limits_{i \in \mathcal{V}} {\sum\limits_{t = 1}^n {c_i^t} }$, ${t_{V2I}} \le n$. Now, we analyze the constraints of the V2I phase.

First, vehicle $i$ can communicate with the RSU only within the RSU's coverage. Therefore, we can obtain the following constraint
\begin{equation}
c_i^t = 0,\forall i \in \mathcal{V},\forall t \in \mathcal{T}, {\rm If} \;{d_i} > {R_r}.
\label{eq8}
\end{equation}
Second, all V2I links should meet their QoS requirements, which can be expressed as follows
\begin{equation}
SN{R_i} \ge t{h_{min}},\forall i \in \mathcal{V}, {\rm If} \;c_i^t = {1},\forall t \in \mathcal{T}.
\label{eq9}
\end{equation}
Third, the RSU can only communicate with one vehicle in any time slot, which can be expressed as follows
\begin{equation}
\sum\limits_{i = 1}^N {c_i^t \le 1,} \forall t \in \mathcal{T}.
\label{eq10}
\end{equation}
Finally, the amount of data received by the vehicle completing the content downloading should be greater than or equal to the content size. Therefore, we can obtain the following constraint
\begin{equation}
\sum\limits_{t = 1}^n {(c_i^t \cdot R_i^t \cdot \Delta )}  \ge D,\forall i \in \mathcal{V}.
\label{eq11}
\end{equation}

In the V2V phase, FD communications and concurrent transmissions are exploited to improve transmission efficiency. We denote the transmission schedule in the V2V transmission phase by $\textbf{S}$, which has $K$ pairings \cite{Niu1}. For each vehicle $i \in (\mathcal{V} - {\mathcal{V}_{V2I}})$, we denote its downloading source by $s_i$, and $s_i$ is the vehicle that has completed the content downloading. We define links ${l_{{s_i},i}}$ and ${l_{{s_j},j}}$ to be adjacent if links ${l_{{s_i},i}}$ and ${l_{{s_j},j}}$ have shared nodes i.e., ${s_i} = {s_j}$ or $i = j$. Adjacent links cannot be scheduled concurrently. To reduce the number of adjacent links for better usage of spatial reuse in transmission scheduling, each vehicle is allowed to be the downloading source of one nearby vehicle once. For each pairing, we define an $N \times N$ matrix $\textbf{A}^k$ to indicate the links scheduled to communicate in the $k$th pairing. The rows of $\textbf{A}^k$ indicate the downloading sources of vehicles, while the columns of $\textbf{A}^k$ indicate all vehicles that need to receive the content. The $(i,j)$th element of $\textbf{A}^k$, $a_{i,j}^k$ indicates whether link $(i,j)$ is scheduled for transmission in the $k$th pairing. If link $(i,j)$ is scheduled for transmission in the $k$th pairing, $a_{i,j}^k = 1$; otherwise, $a_{i,j}^k =0$. We denote the number of time slots for completing the $k$th pairing schedule by $\delta^k$. To maximize transmission efficiency, the transmission schedule should complete the content downloading for all vehicles with a minimum number of time slots. Therefore, the number of time slots for vehicles, having not downloaded content in the V2I phase, to complete the content sharing in the V2V phase is
${t_{V2V}} = \sum\limits_{k = 1}^K {{\delta ^k}}$, ${t_{V2V}} < n$. Now, we analyze the constraints of the V2V phase.

First, in the schedule, all vehicles can only be the download source once, and vehicles $i$ without downloading content can only receive the content transmission of one vehicle. Therefore, we can obtain the following constraint,
\begin{equation}
\sum\limits_{k = 1}^K {a_{{s_i},i}^k = 1,} \forall i \in (\mathcal{V} - {\mathcal{V}_{V2I}}),\forall {s_i} \in \mathcal{V}.
\label{eq12}
\end{equation}

Second, since vehicle $i$ obtains the common content after vehicle $s_i$ received the common content, the downloading of vehicle $s_i$ should be scheduled ahead of the downloading for vehicle $i$, which can be formulated as follows. This constraint represents a group of constraints since $\widetilde K$ varies from 1 to $K$, i.e.,
\begin{equation}
\resizebox{.98\hsize}{!}{$\sum\limits_{k = 1}^{\widetilde K} {a_{{s_{{s_i}}},{s_i}}^k}\! \ge\! \sum\limits_{k = 1}^{\widetilde K} {a_{{s_i},i}^k,} \forall i\! \in\! (\mathcal{V}\! -\! {\mathcal{V}_{V2I}}),\widetilde K\! =\! 1\!\thicksim \!K, {\rm If} \; \forall {s_i}\! \in\! (\mathcal{V}\! -\! {\mathcal{V}_{V2I}}).$}
\label{eq13}
\end{equation}

Third, FD communications are exploited in the V2V phase, and each transmission link can only carry out two-hop communications at most, which can be expressed as follows
\begin{equation}
a_{{s_{{s_i}}},{s_i}}^k + a_{{s_i},i}^k \le 2,\forall k,\forall i \in (\mathcal{V} - {\mathcal{V}_{V2I}}).
\label{eq14}
\end{equation}

Finally, the content downloading of the vehicles that have not downloaded content in the V2I phase should be completed. We can obtain the related constraint as follows
\begin{equation}
\sum\limits_{k = 1}^K {(a_{{s_i},i}^k \cdot \sum\limits_{t = {t_{V2I}} + \sum\limits_{k = 1}^{K - 1} {{\delta ^k}}  + 1}^{{t_{V2I}} + \sum\limits_{k = 1}^K {{\delta ^k}} } {R_{{s_i},i}^k}  \cdot \Delta) \ge D,} \forall i \in (\mathcal{V} - {\mathcal{V}_{V2I}}).
\label{eq15}
\end{equation}

In summary, the problem of optimal transmission scheduling can be formulated as follows.
\begin{equation}
\min ({t_{V2I}} + {t_{V2V}}) = \min (\sum\limits_{i \in \mathcal{V}} {\sum\limits_{t = 1}^n {c_i^t + } } \sum\limits_{k = 1}^K {{\delta ^k}} ),
\label{eq16}
\end{equation}
\hspace{2cm}s.t.
\hspace{0.2cm}Constraints (\ref{eq8}) -- (\ref{eq15}).

\noindent This optimal problem is a nonlinear integer programming problem, and is NP-hard \cite{Relay}. We propose the selection approach to minimize the number of time slots for content distribution in the next section. Besides, we define the transmission utility function for selecting vehicles to receive the content from the RSU in the V2I phase.

\begin{algorithm}[t]
\caption{V2I Transmission Path Selection Algorithm}
\label{alg1}
\begin{algorithmic}[1]
\STATE \textbf{Initialization:} Obtain the location of each vehicle;\\
 $t_{V2I} = 0$; $m_i = 0$; $D_i^t = 0$; $\mathcal{V}_t = \emptyset$; $\mathcal{V}_a = \emptyset$; $\mathcal{V}_b = \mathcal{V}$;
\WHILE{${V_N} \in {\mathcal{V}_b}$}
\FOR{vehicle $i$ $(1 \le i \le N)$}
\STATE $t = {t_{V2I}} + 1$;
\IF{vehicle $i$ $(i \in {\mathcal{V}_b})$}
\IF{$d_i^t \le {R_r}$}
\STATE ${m_i} = {m_i} + 1$; $D_i^t = D_i^t + R_i^t \cdot \Delta$;
\IF{$D_i^t < D$}
\STATE $t = t + 1$;
\STATE Go back to line 6;
\ELSE
\STATE Obtain vehicle $j$ $(j \in {\mathcal{V}_b}\backslash i)$ with the smallest number of time slots $m_{i,j}$ from vehicle $i$;
\STATE Obtain vehicle $g$ $(g \in {\mathcal{V}_b}\backslash \{ i,j\})$ with the smallest number of time slots $m_{j,g}$ from vehicle $j$;
\STATE ${m_i^\prime} = \max ({m_{i,j}},{m_{j,g}})$;
\STATE ${\mathcal{V}_t} = {\mathcal{V}_t} \cup \{ i\}$;
\STATE ${U_i} = \{ {m_i} + {m_i^\prime}\}$;
\ENDIF
\ENDIF
\ENDIF
\ENDFOR
\STATE ${\mathcal{V}_a} = {\mathcal{V}_a} \cup \{ \arg \mathop {\min }\limits_{i \in {\mathcal{V}_t}} \{ {U_i}\} \}$; ${\mathcal{V}_b} = {\mathcal{V}_b} - \{ \arg \mathop {\min }\limits_{i \in {\mathcal{V}_t}} \{ {U_i}\} \}$;
\STATE ${t_{V2I}} = {t_{V2I}} + {m_{\arg \mathop {\min }\limits_{i \in {\mathcal{V}_t}} \{ {U_i}\} }}$;
\ENDWHILE
\STATE \textbf{Return} $\mathcal{V}_a$, $\mathcal{V}_b$ and $t_{V2I}$.
\end{algorithmic}
\end{algorithm}

\section{Joint V2I and V2V scheduling algorithm}\label{S5}

\subsection{V2I Transmission Path Selection Algorithm}\label{S5-1}

In content distribution scheduling, to effectively exploit content forwarding in the V2V phase, the RSU should select the appropriate vehicles within its coverage to download the integrity content based on the vehicular network topology and the distance between the RSU and vehicles. Therefore, the priority rules for the RSU to select appropriate vehicles are as follows.

1) Within the RSU's coverage, vehicles with fewer time slots to complete the integrity content downloading have priority to receive the content from the RSU.

2) In the V2V phase, each vehicle can receive content and send the content to other vehicles simultaneously. Within the RSU's coverage, vehicles with fewer time slots to complete the content downloading in two-hop FD communications of the V2V phase have a higher priority.

Based on the priority description above, we define the transmission utility function of the vehicle with the highest priority within the RSU's coverage at a given scheduling time slot $t$ as.
\begin{equation}
\mathop {\min }\limits_{i \in {\mathcal{V}_t}} {U_i} = \mathop {\min }\limits_{i \in {\mathcal{V}_t}} \{ {m_i} + {m_i^\prime}\} ,
\label{eq17}
\end{equation}
where, $\mathcal{V}_t$ denotes the set of vehicles that are within the coverage of the RSU and can complete the content downloading in time slot $t$, $m_i$ denotes the number of time slots in which vehicle $i$ completes the content downloading in the V2I phase in time slot $t$, and $m_i^\prime$ denotes the maximum number of time slots for vehicle $i$ to complete the content downloading in two-hop FD communications.

The pseudo-code of the V2I transmission path selection algorithm is presented in Algorithm \ref{alg1}. The algorithm obtains the location of each vehicle. The set of vehicles that have completed the content downloading are not selected as downloading sources is denoted by $\mathcal{V}_a$, and the set of vehicles that do not have the content downloaded is denoted by $\mathcal{V}_b$. Lines 5--21 select a vehicle with the smallest transmission utility function to communicate with the RSU according to (\ref{eq16}) within RSU's coverage. Lines 6--10 calculate the number of time slots in which vehicle $i \in {\mathcal{V}_b}$ within the coverage of the RSU completes the content downloading in time slot $t$. Line 12 obtains vehicle $j \in {\mathcal{V}_b}\backslash i$ with the smallest number of time slots from vehicle $i$, and line 13 obtains vehicle $g \in {\mathcal{V}_b}\backslash \{ i,j\}$ with the smallest number of time slots from vehicle $j$. Thus, the maximum number of time slots for vehicle $i$ to complete the content downloading in two-hop FD communications, as in line 14. Line 22 calculates the number of time slots $t_{V2I}$ for content distribution in the V2I phase. Finally, we obtain the set of vehicles that have completed the content downloading and are not selected as downloading sources, the set of vehicles that do not have the content downloaded, and the number of time slots for content distribution in the V2I phase.

\subsection{V2V Full-duplex Concurrent Transmission Scheduling Algorithm}\label{S5-2}

\begin{algorithm}[!t]
\caption{V2V Full-duplex Concurrent Transmission Scheduling Algorithm}
\label{alg2}
\begin{algorithmic}[1]
\STATE \textbf{Input:} $\mathcal{V}_a$; $\mathcal{V}_b$; $t_{V2I}$;
\STATE \textbf{Initialization:} $k = 0$; $t = t_{V2I}$; $D_{i,j}^t = 0$; $t_{V2V} = 0$;
\WHILE{$\left| {{\mathcal{V}_b}} \right| > 0$}
\STATE $k = k + 1$; ${\delta ^k} = 0$; ${S^k} = \emptyset$; $S_s^k = \emptyset$;
\FOR{vehicle $i$ $(i \in {\mathcal{V}_a})$}
\STATE  Obtain vehicle $j$ $(j \in {\mathcal{V}_b})$ with the largest transmission rate link $l_{i,j}$ from vehicle $i$;
\STATE ${S^k} = {S^k} \cup \{ {l_{i,j}}\}$;
\ENDFOR
\STATE Calculate the number of time slots ${\xi _{{l_{i,j}}}}$ of each link in $S^k$, and sort these links by the number of time slots from small to large;
\STATE $S_s^k = S_s^k \cup \{ \arg \mathop {\min }\limits_{{l_{i,j}} \in {S^k}} \{ {\xi _{{l_{i,j}}}}\} \}$;\\ ${S^k} = {S^k} - \{ \arg \mathop {\min }\limits_{{l_{i,j}} \in {S^k}} \{ {\xi _{{l_{i,j}}}}\} \}$;
\STATE Obtain vehicle $g$ $(g \in {\mathcal{V}_b})$ with the largest transmission rate link $l_{j,g}$ from vehicle $j$;
\IF {link $l_{j,g}$ does not conflict with the link in $S_s^k$}
\STATE $S_s^k = S_s^k \cup {l_{j,g}}$; ${\mathcal{V}_a} = {\mathcal{V}_a} - \{ j\}  \cup \{ g\}$; ${\mathcal{V}_b} = {\mathcal{V}_b} - \{ g\}$;
\ENDIF
\FOR {link $l_{i,j}$ $({l_{i,j}} \in {S^k})$}
\IF {link $l_{i,j}$ does not conflict with the link in $S_s^k$}
\STATE $S_s^k = S_s^k \cup {l_{i,j}}$; ${\mathcal{V}_a} = {\mathcal{V}_a} - \{ i\}  \cup \{ j\}$; ${\mathcal{V}_b} = {\mathcal{V}_b} - \{ j\}$;
\STATE Obtain vehicle $g$ $(g \in {\mathcal{V}_b})$ with the largest transmission rate link $l_{j,g}$ from vehicle $j$;
\IF {link $l_{j,g}$ does not conflict with the link in $S_s^k$}
\STATE $S_s^k = S_s^k \cup {l_{j,g}}$; ${\mathcal{V}_a} = {\mathcal{V}_a} - \{ j\}  \cup \{ g\}$;\\
 ${\mathcal{V}_b} = {\mathcal{V}_b} - \{ g\}$;
\ENDIF
\ENDIF
\ENDFOR
\WHILE{$\left| {S_s^k} \right| > 0$}
\STATE $t = t + 1$;
\FOR{link $l_{i,j}$ $({l_{i,j}} \in S_s^k)$}
\STATE Calculate the actual transmission rate $R_{i,j}^t$ of link $l_{i,j}$;
\STATE ${m_{i,j}} = {m_{i,j}} + 1$; $D_{i,j}^t = D_{i,j}^t + R_{i,j}^t \cdot \Delta$;
\IF{$D_{i,j}^t > D$}
\STATE $S_s^k = S_s^k - {l_{i,j}}$;
\IF{${m_{i,j}} > {\delta ^k}$}
\STATE ${\delta ^k} = {m_{i,j}}$;
\ENDIF
\ENDIF
\ENDFOR
\ENDWHILE
\STATE ${t_{V2V}} = {t_{V2V}} + {\delta ^k}$;
\ENDWHILE
\STATE \textbf{Return} $t_{V2V}$.
\end{algorithmic}
\end{algorithm}

When multiple vehicles send the same content downloading request to the RSU, the RSU selects some vehicles to access the Internet according to Algorithm \ref{alg1} with completing the content downloading. Then other vehicles share the content through V2V communications. The pseudo-code of the full-duplex concurrent transmission scheduling algorithm in the V2V phase is presented in Algorithm \ref{alg2}. The algorithm iteratively schedules the vehicles in $\mathcal{V}_b$ into the transmission paths until all vehicles are scheduled, as in line 3. After obtaining $\mathcal{V}_a$ and $\mathcal{V}_b$ from Algorithm \ref{alg1}, we obtain the pre-scheduled set $S^k$ of the first-hop communication link in the $k$th scheduled transmission path, as in lines 5--8. Lines 9--23 determine the link scheduling set $S_s^k$ in the $k$th scheduling according to the contention graph. If the vehicle has been selected as the downloading source, it will be removed from $\mathcal{V}$, and if the vehicle has downloaded the content, it will be removed from $\mathcal{V}_b$, as in line 20. From lines 24 to 36, the algorithm obtains the number of time slots needed for each flow in $S_s^k$, and obtain the number of time slots needed for the $k$th pairing, $\delta^k$. In line 27, the algorithm obtains the transmission rate of flow $l_{i,j}$ scheduled in the $k$th pairing, $R_{i,j}^t$, and its needed number of time slots is $m_{i,j}$. In lines 31--32, $\delta^k$ is updated to accommodate the time slot demands of all flows in the $k$th pairing. Finally, the scheduling results $t_{V2V}$ is outputted in line 39.

\subsection{Complexity Analysis}\label{S5-3}

The content distribution scheme based on joint V2I and V2V scheduling includes two algorithms. In the V2I phase, when $N$ vehicles request content downloading from the RSU, the RSU needs to judge the location of vehicles and select the transmission path through Algorithm \ref{alg1}. For the computational complexity of Algorithm \ref{alg1}, the outer while loop in line 2 has $\left| {{\mathcal{V}_b}} \right|$ iterations, which at most is $N$, and the inner for loop in line 3 has $N$ iterations. Besides, the complexity of the minimum number of time slots for the first hop and the second hop communication is obtained in the three-node FD transmission, which at most are $N-2$ and $N-3$, respectively, as in lines 12-13. Thus, the computational complexity of Algorithm \ref{alg1} is ${\rm \mathcal{O}}((N - 2){N^2})$.
In the V2V phase, FD communications and concurrent transmissions are exploited to download the content among vehicles. For the computational complexity of Algorithm \ref{alg2}, since the outer while loop in line 3 has $\left| {{\mathcal{V}_b}} \right|$ iterations, which at most is $N-1$. The for loop in line 15 has $\left| {S^k} \right|$ iterations, which at most is $N-2$, line 16 judges whether link $l_{i,j}$ conflicts with the link in $\left| { S_s^k } \right|$, which has $N-1$ iterations, and line 19 judges whether link $l_{j,g}$ conflicts with the link in $\left| { S_s^k } \right|$, which has $N-1$ iterations. Thus, the computational complexity of Algorithm \ref{alg2} is ${\rm \mathcal{O} ({(N - 2)(N - 1)^3})}$.
Therefore, combined with Algorithm \ref{alg1} and \ref{alg2}, the complexity of the two-phase algorithm is ${\rm \mathcal{O} ({(N - 2)(N - 1)^3})}$.

\section{SIMULATION RESULTS}\label{S6}

\subsection{Simulation Setup}\label{S6-1}

In the simulation, we consider a straight five-lane highway which has been introduced in Section~\ref{S3}. The width of each lane is 4 meters. The length of the road is 2000 m and the RSU is located at 500 m away from the left end. 100 vehicles are injected from the left end by following a Poisson distribution with $\lambda$ vehicles per second, and the speed of the vehicles is 20 m/s. The content size is 3 Gb. The communication range of RSU is 200 m. The communication range of the vehicle is 200 m in V2I communications and 20 m in V2V communications. This is because each connection builds in different communication modes, including infrastructure mode and vehicular ad hoc mode. The connection in vehicular ad hoc mode is unstable because of the high mobility of the vehicles. Thus, we assume the vehicles cannot directly communicate with others when the distance between them is long as more than 20 m. We assume that all vehicles cannot change lanes. V2I and V2V communications are in a 28 GHz mmWave vehicular network. To ensure that content can be transmitted normally when vehicles are moving fast, we update the positions of vehicles and calculate new transmission rates of links every slot, and then the transmission scheme continues to be executed. The simulations are built in MATLAB R2019b platform and parameters are listed in Table \ref{table1}.

\begin{table}[t]
\begin{center}
\caption{SIMULATION PARAMETERS}
\begin{tabular}{ccc}
\hline
Parameter & Symbol  & Value \\
\hline
Number of vehicles & $N$ & 100 \\
Speed of vehicle & $v$ & 20 m/s \\
Communication range of RSU & $R_r$ & 200 m \\
Communication range of vehicle & $R$ & 20 m \\
Content size & $D$ & 3 Gb \\
Transmission power of RSU & $P_t$ & 30 dBm \\
Transmission power of vehicle & $P_v$ & 20 dBm \\
System bandwidth & $W$ & 800 MHz \\
Background noise & $N_0$ & -134 dBm/MHz \\
SINR threshold of the V2I and V2V link & ${th}_{min}$ & 20 dB \\
SI cancelation level & $\beta$ & ${10}^{-8}$ \\
Path loss exponent & $\tau$ & 2 \\
MUI factor & $\rho$ & 1 \\
Slot duration & $\Delta$ & 0.1 ms \\
Number of time slots in transmission phase & $\left| \mathcal{T} \right|$ & $10^6$ \\
The half-power beam width of antenna & $\phi$ & $30^\circ$ \\
Sidelobe gain & $g$ & 0.1 \\
\hline
\end{tabular}
\label{table1}
\end{center}
\end{table}

To show the advantages of our scheme in terms of transmission time slots, throughput, and energy consumption, we evaluate the following three performance metrics:

1) \emph{Transmission Time Slots:} The number of transmission time slots for all requested vehicles to complete the content downloading.

2) \emph{System Throughput:} The achieved throughput of all requested vehicles to complete the content downloading in mmWave vehicular networks, which can be expressed as
\begin{equation}
ST = \frac{{\left| \mathcal{V} \right| \cdot D}}{{({t_{V2I}} + {t_{V2V}}) \cdot \Delta}},
\label{eq18}
\end{equation}
where ST represents ``System Throughput''.

3) \emph{Energy Consumption:} The total energy consumption of transmissions in mmWave vehicular networks, which can be expressed as
\begin{equation}
EC = {t_{V2I}} \cdot \Delta \cdot {P_t} + \sum\limits_{k = 1}^K {\sum\limits_{w \in {\mathcal{V}^k}} {\frac{D}{{R_w^k}} \cdot {P_v}} } ,
\label{eq19}
\end{equation}
where we denote the transmission rate of the flow $w$ in the $k$th pairing by ${R_w^k}$, and EC represents ``Energy Consumption''.

In the simulation, our scheme is compared with the following three schemes:

1) \emph{FCFS cooperation scheme:} The most commonly used scheduling approach for file transmission is the FCFS scheme, which is a first-come, first-served mechanism without considering other factors in the scheduling process. In the V2I phase, the RSU communicates with the vehicle that drives into its coverage area first. In the V2V phase, FD communications and concurrent transmissions are exploited to download the content among vehicles.

2) \emph{Random cooperation scheme:} In the V2I phase, the RSU randomly selects vehicles within its coverage to communicate. In the V2V phase, vehicles that have completed the content downloading randomly communicates with vehicles that have not downloaded the content.

3) \emph{Non-cooperative scheme:} In content distribution scheduling, we only consider the V2I phase, not the V2V phase. The RSU selects the vehicles with the shortest distance within its coverage to communicate, and there are no cooperative communications between vehicles.

\subsection{Simulation Results}\label{S6-2}

\begin{figure}[t]
\begin{center}
\includegraphics*[width=0.85\columnwidth,height=2.5in]{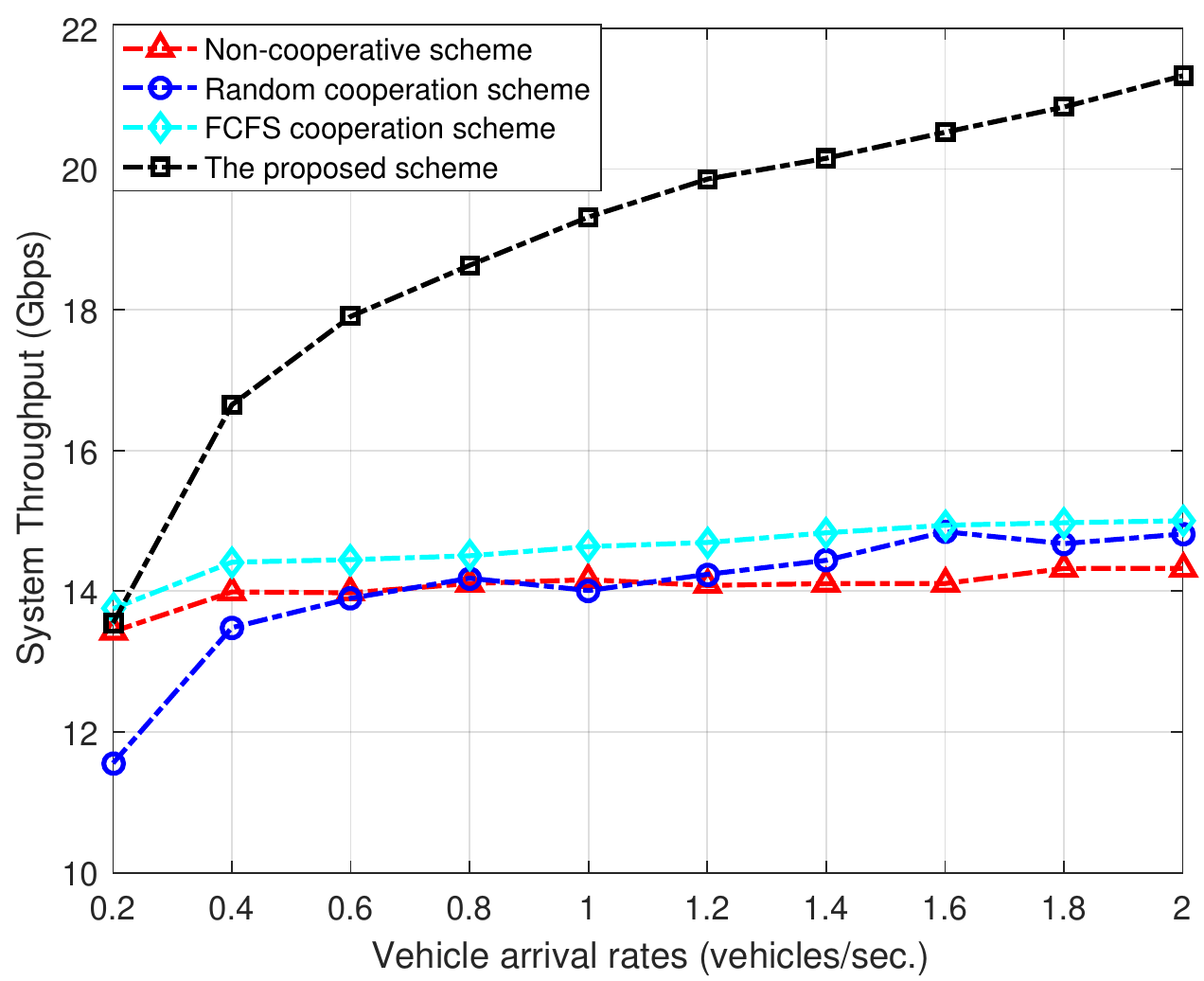}
\end{center}
\caption{System throughput of the three transmission schemes under different vehicle arrival rates.}
\label{fig3}
\end{figure}

\begin{figure}[t]
\begin{center}
\includegraphics*[width=0.85\columnwidth,height=2.5in]{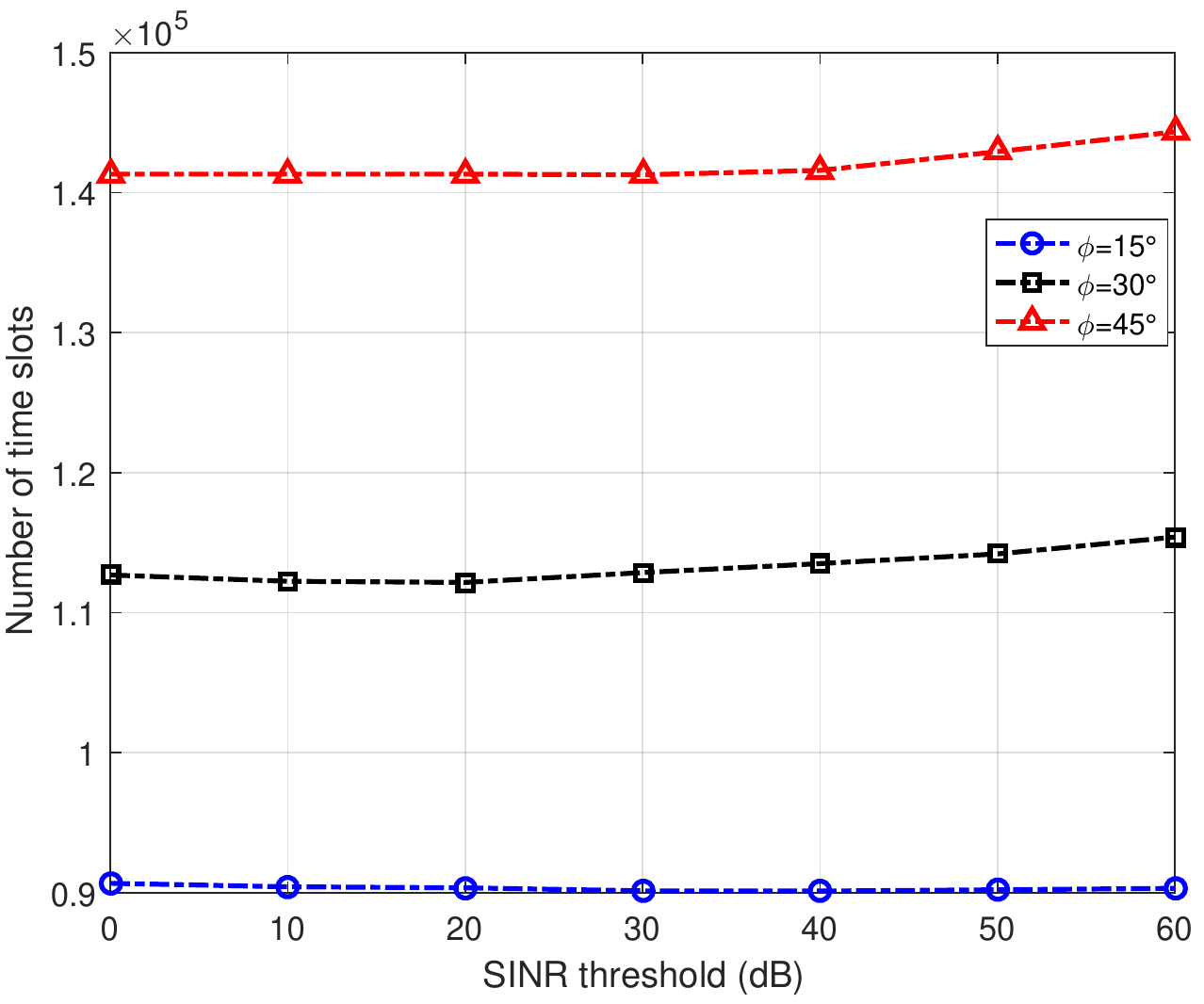}
\end{center}
\caption{Number of time slots of the proposed scheme with different SINR thresholds and beam width.}
\label{fig4}
\end{figure}

In Fig. \ref{fig3}, we plot the system throughput of the three transmission schemes under different vehicle arrival rates. We can see that the system throughput of four transmission schemes is increasing with the increased arrival rate of vehicles. This is because a higher vehicle arrival rate leads to higher connectivity which is critical for multi-hop V2V communications. Specifically, thanks to the higher vehicle arrival rate, the inter-distances between vehicles are smaller, which is beneficial for RSU to select the appropriate vehicles for communication. When the vehicle arrival rate is 2 vehicles per second, compared with the random cooperation scheme and FCFS cooperation scheme, the proposed scheme increases the system throughput by 43.9$\%$ and 42.1$\%$, respectively. The proposed scheme achieves better performance in mmWave vehicular networks with a high vehicle arrival rate. Therefore, in remaining simulations of this paper, we set the value of the vehicle arrival rate to be 2 vehicles per second.

In Fig. \ref{fig4}, we plot the number of time slots of the proposed scheme with different SINR thresholds and beam width. With the increase of the SINR threshold, the time slots numbers of the scheme decreases first and then increases. Properly increasing of the SINR threshold can avoid some transmissions with very small rates. The rates of other transmissions can be increased, which will lead to shorter transmit times. When the SINR threshold is increased to a certain value (i.e., $\phi=30^\circ$ and $th_{min}$ = 20 dB in Fig. \ref{fig4}), the number of time slots of the scheme starts to increase. This is because that too small SINR thresholds weaken the advantage of spatial reuse and increases the number of content distribution time slots. So an appropriate actual SINR threshold is important for transmission performances. In other simulations of this paper, we set the value of the SINR threshold to be 20 dB.

\begin{figure}[t]
\begin{center}
\includegraphics*[width=0.85\columnwidth,height=2.5in]{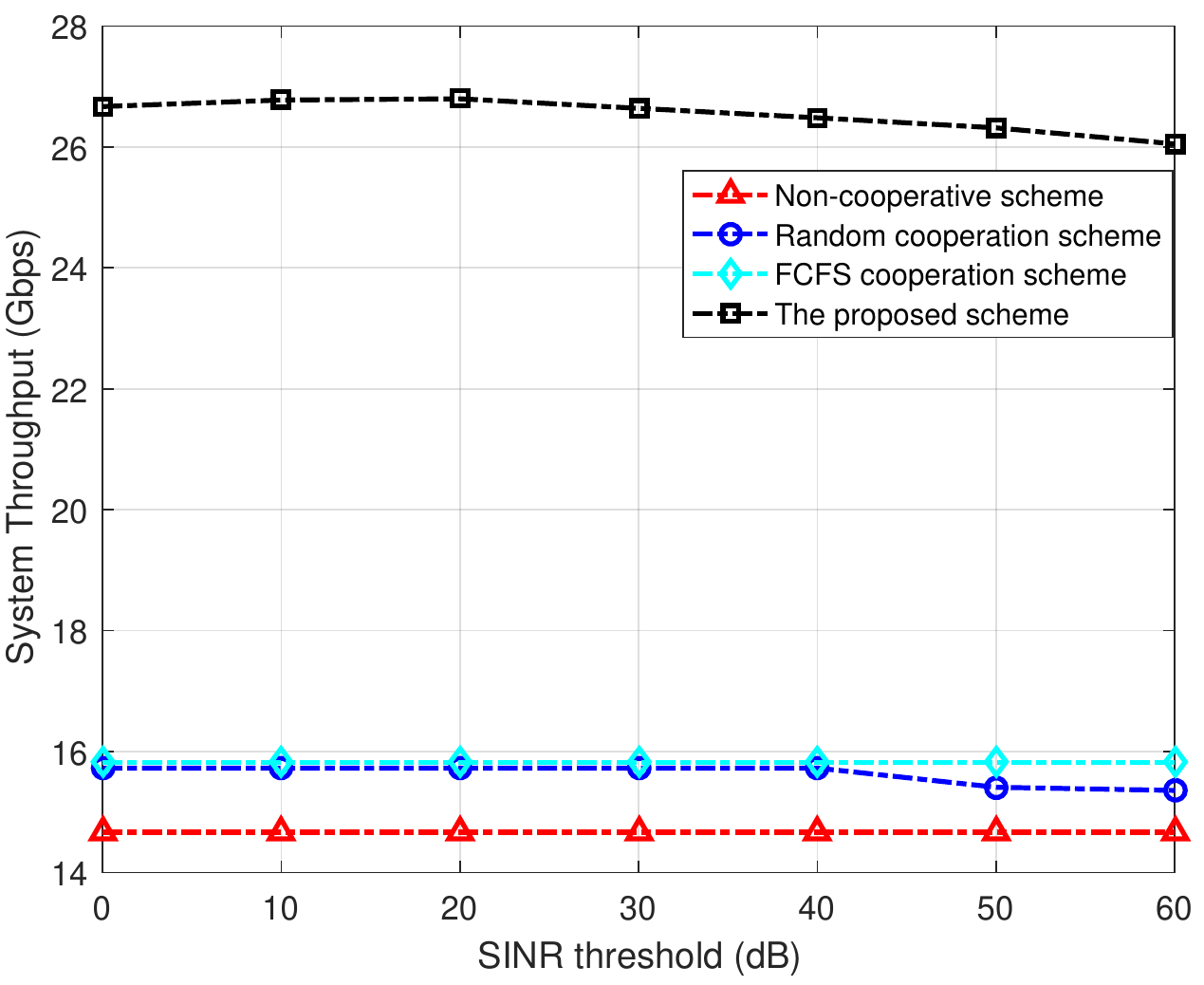}
\end{center}
\caption{System throughput versus the SINR threshold.}
\label{fig5}
\end{figure}

\begin{figure}[t]
\begin{center}
\includegraphics*[width=0.85\columnwidth,height=2.5in]{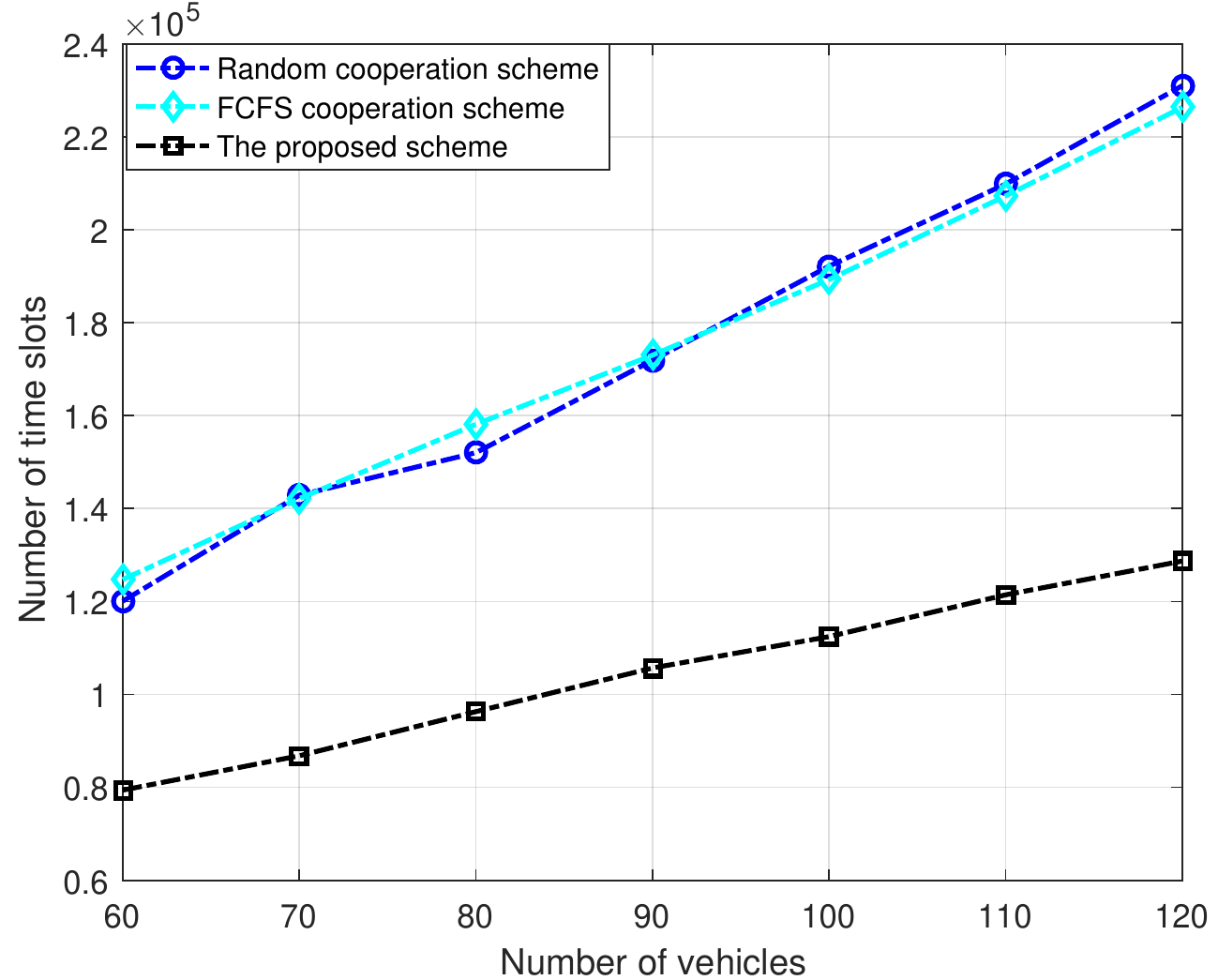}
\end{center}
\caption{Number of time slots versus the number of vehicles.}
\label{fig6}
\end{figure}

In Fig. \ref{fig5}, we plot the system throughput under different SINR thresholds. We can see that with the increase of the SINR threshold, the system throughput of the non-cooperative scheme and FCFS cooperation scheme remains unchanged, while the system throughput of other schemes decreases. This is because the RSU exploits serial communication in the V2I phase, and there is only a LOS link between the RSU and vehicles. The SINR threshold does not affect V2I communications but affects V2V communications. Therefore, the system throughput of the random cooperation scheme and the proposed scheme changes little with the increase of the SINR threshold.

In Fig. \ref{fig6}, we plot the number of content distribution time slots under different vehicle numbers. The non-cooperative scheme cannot guarantee that all requested vehicles complete content downloading, and so we present only three result curves. We can see that the number of time slots in the proposed scheme is significantly less than the random cooperative scheme as the number of vehicles increases. When the number of vehicles is 100, compared with the random cooperation scheme and FCFS cooperation scheme, the proposed scheme reduces the transmission time by about 41.5$\%$ and 40.6$\%$, respectively. The reason for the proposed scheme to achieve fewer transmission slots is that when the RSU selects vehicles to download the content, it fully considers the vehicular network topology and the transmission scheduling between vehicles.

\begin{figure}[t]
\begin{center}
\includegraphics*[width=0.85\columnwidth,height=2.5in]{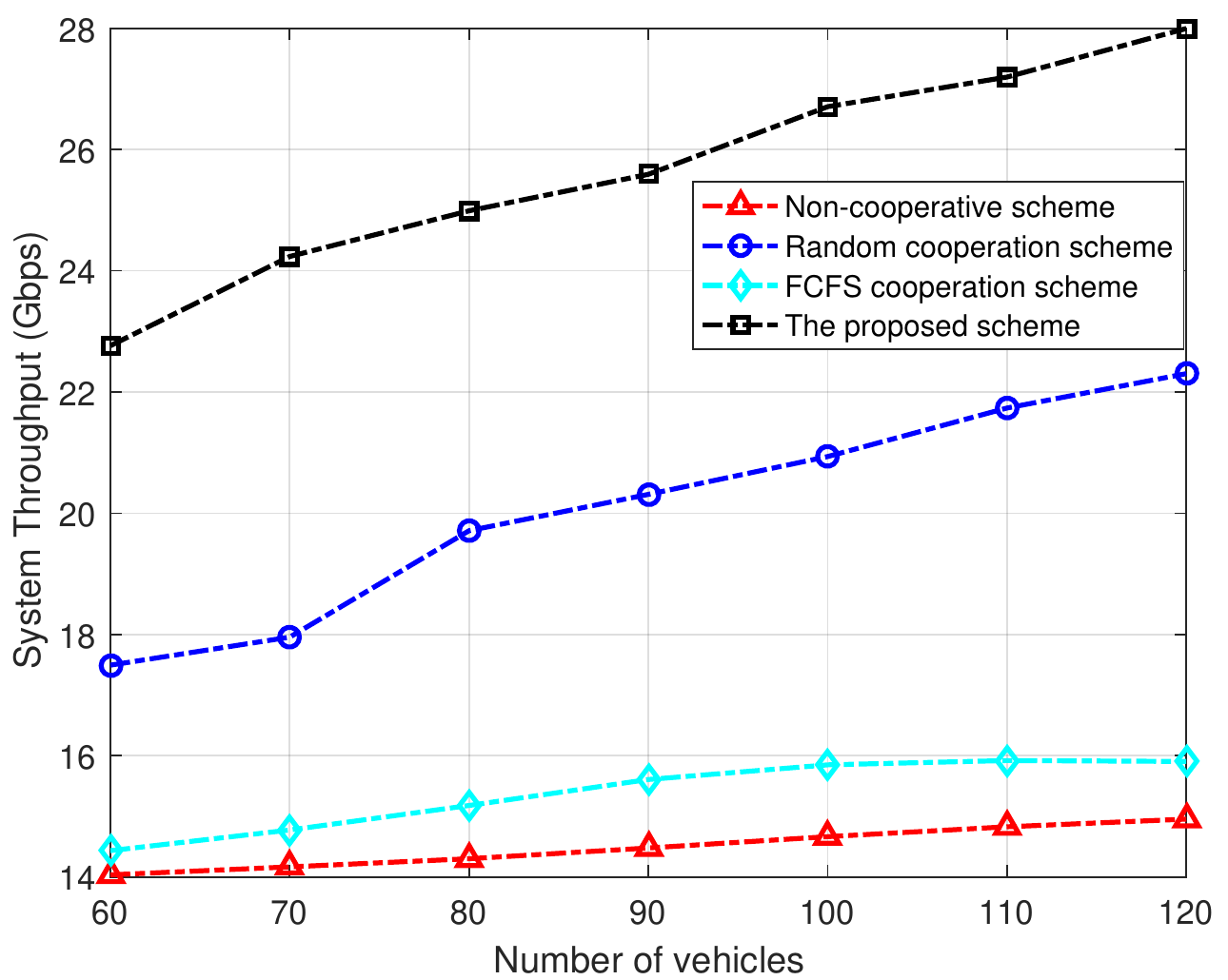}
\end{center}
\caption{System throughput versus the number of vehicles.}
\label{fig7}
\end{figure}

\begin{figure}[t]
\begin{center}
\includegraphics*[width=0.85\columnwidth,height=2.5in]{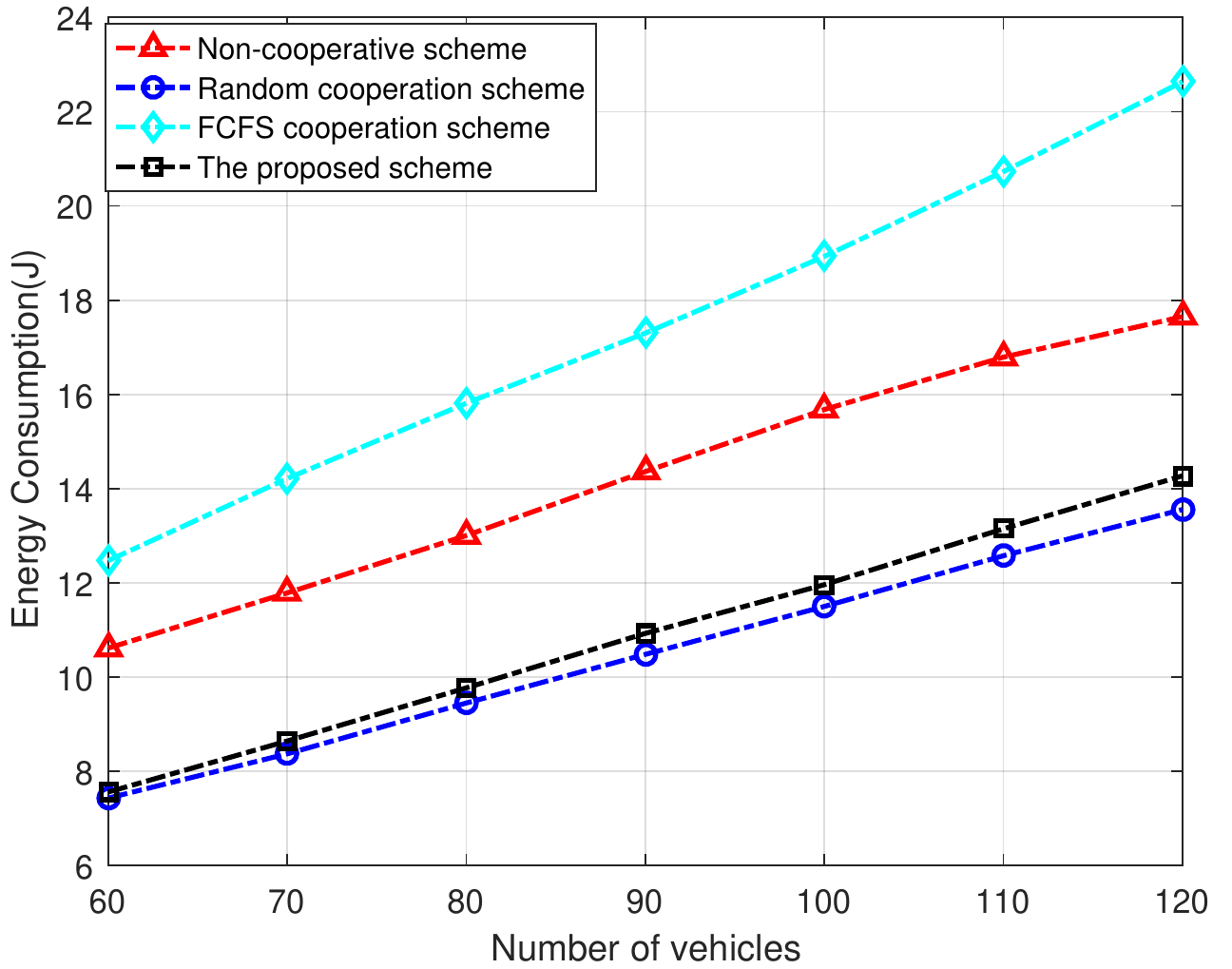}
\end{center}
\caption{Energy Consumption versus the number of vehicles.}
\label{fig8}
\end{figure}

In Fig. \ref{fig7}, we plot the system throughput under different numbers of vehicles. We can see that the system throughput of the four schemes is increasing with the increased number of vehicles. For the non-cooperative scheme and FCFS cooperation scheme, the trend of system throughput is rising slowly. This is because, in the non-cooperative scheme, vehicles only download the content in the V2I phase and without V2V communications after vehicles leave the coverage of the RSU. Therefore, only some vehicles have completed the content downloading. In the FCFS cooperation scheme, vehicles also mainly download content in the V2I phase. The gap between the proposed scheme and the non-cooperative scheme is even larger because of the joint V2I and V2V scheduling, V2V communications in close proximity, and concurrent transmissions are not exploited in the non-cooperative scheme.

\begin{figure}[t]
\begin{center}
\includegraphics*[width=0.85\columnwidth,height=2.5in]{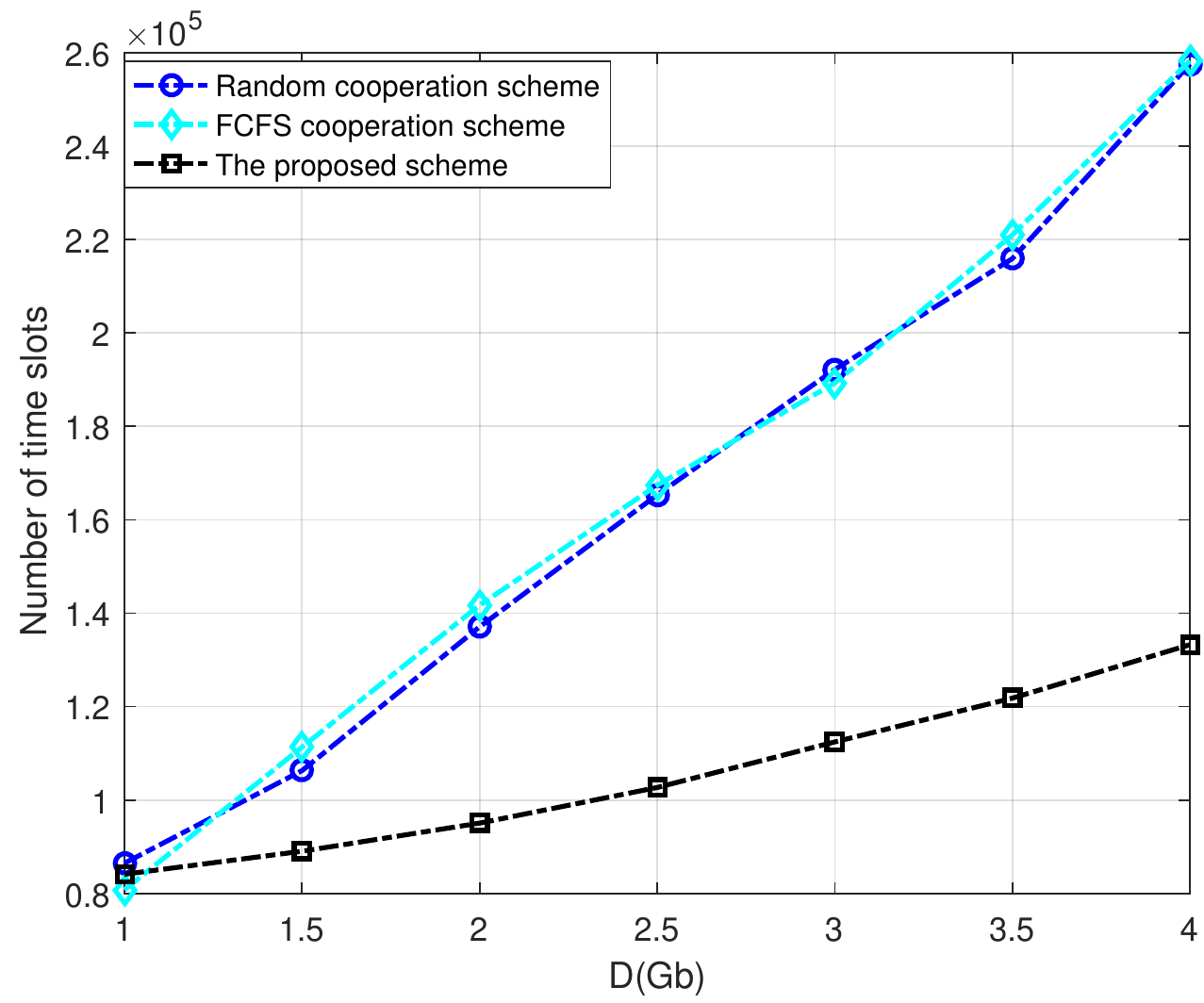}
\end{center}
\caption{Number of time slots versus the content sizes.}
\label{fig9}
\end{figure}

\begin{figure}[t]
\begin{center}
\includegraphics*[width=0.85\columnwidth,height=2.5in]{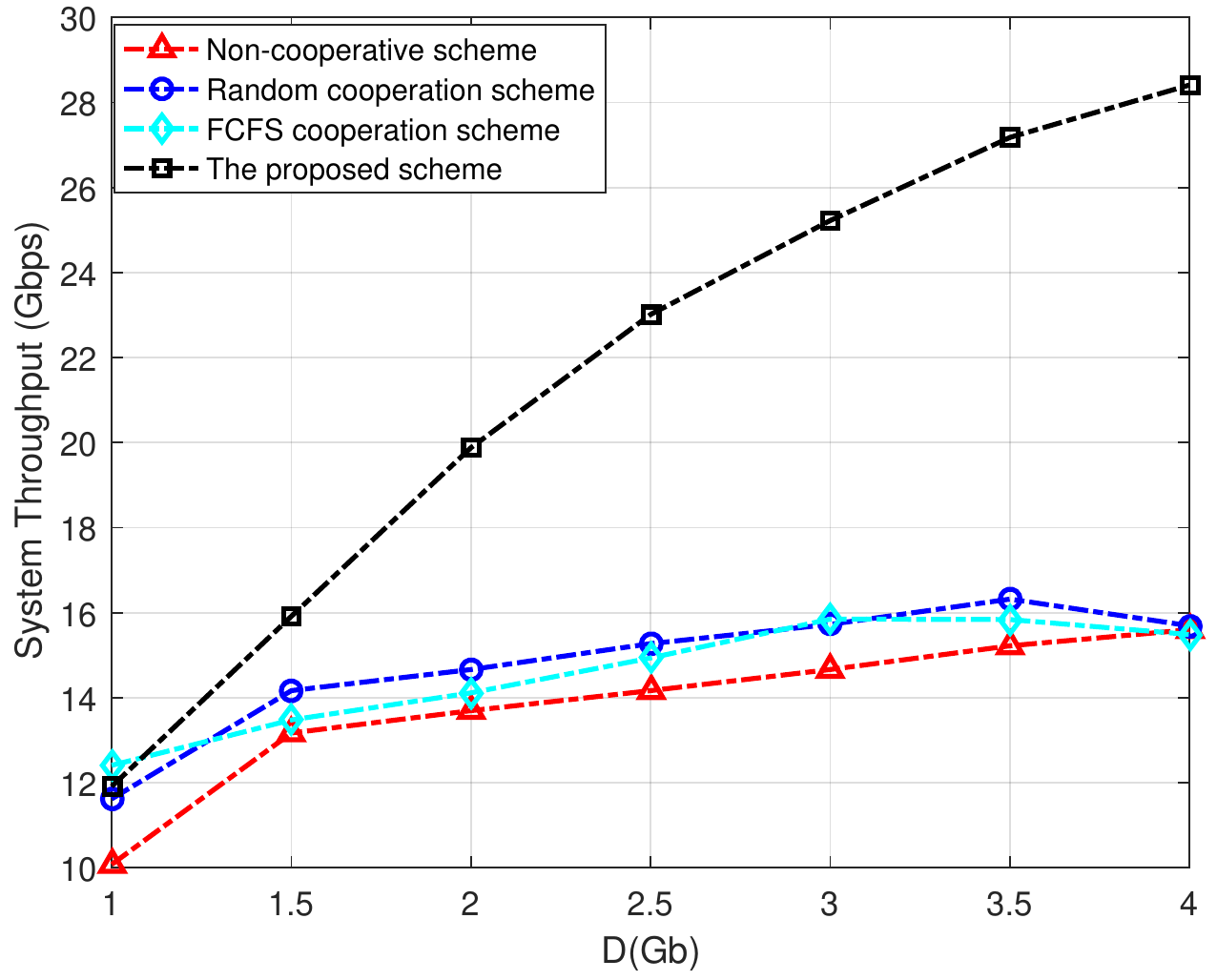}
\end{center}
\caption{System throughput versus the content sizes.}
\label{fig10}
\end{figure}

\begin{figure}[t]
\begin{center}
\includegraphics*[width=0.85\columnwidth,height=2.5in]{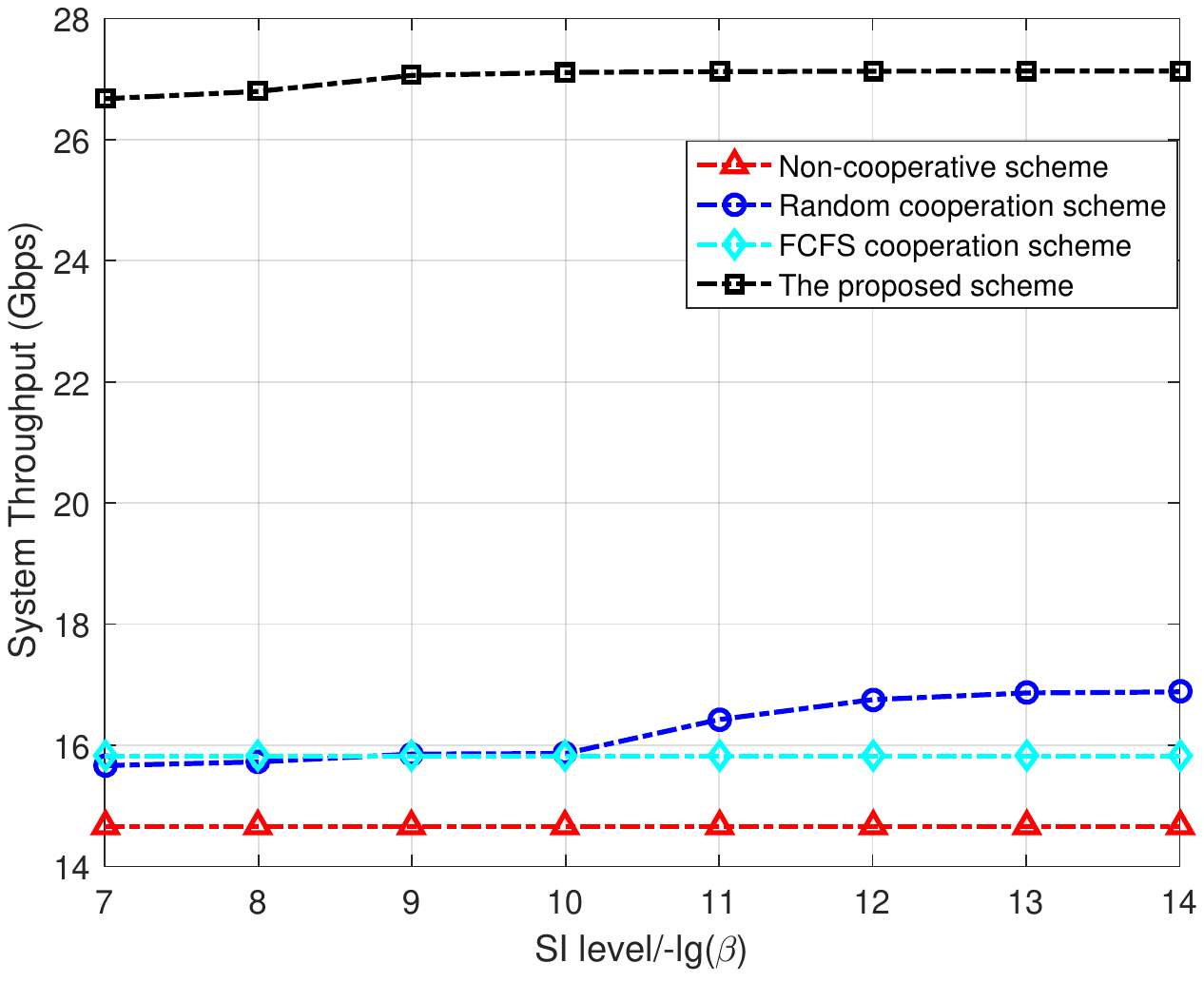}
\end{center}
\caption{System throughput versus the SI level.}
\label{fig11}
\end{figure}

In Fig. \ref{fig8}, we plot the energy consumption of the whole system under different numbers of vehicles. With the increase of vehicle users, energy consumption increases since more vehicle users are needed to be served, and more transmissions are required. As can be seen in the figure, the energy consumption of the FCFS cooperation scheme is the largest. This is because, in the FCFS cooperation scheme, vehicles mainly download content in the V2I phase, and a small number of vehicles utilize V2V communications. In addition, the non-cooperative scheme cannot guarantee that all requested vehicles complete the content downloading, because it only exploits V2I communications without V2V communications. The RSU transmits ten times more power than vehicles. Therefore, the more vehicles that download the content in the V2I phase, the more energy they consume. As can be seen in the figure, the energy consumption of the proposed scheme is more than that of the random cooperation scheme. This is mainly due to the RSU communicating with more vehicles in the proposed scheme.

In Fig. \ref{fig9}, we plot the number of content distribution time slots under different content sizes. The non-cooperative scheme cannot guarantee that all requested vehicles complete the content downloading, so there are only three result curves. We can see that when the content size is small, the gap between the three schemes is small. However, with the increase in content size, the gap between the FCFS cooperation scheme and the proposed scheme is even larger. When the content size is 3 Gb, compared with the random cooperation scheme and FCFS cooperation scheme, the proposed scheme reduces the transmission time by about 41.5$\%$ and 40.6$\%$, respectively.

In Fig. \ref{fig10}, we plot the system throughput under different content sizes. We can observe the gap between the proposed scheme and other schemes. The larger the content size, the better the transmission performance of the proposed scheme. When the content size is 3 Gb, the proposed scheme increases the system throughput by 59.2$\%$ compared with the FCFS cooperation scheme, by 60.4$\%$ compared with the random cooperation scheme, and by 60.9$\%$ compared with the non-cooperative scheme. For downloading a large-size file, the proposed scheme has the best performance. This demonstrates the advantages of joint V2I and V2V scheduling, V2V communications, and concurrent transmissions in the proposed scheme.

In Fig. \ref{fig11}, we plot the system throughput under different SI cancelation levels. The abscissa $x$ is $- \lg (\beta )$, i.e., the SI cancelation level $\beta  = {10^{ - x}}$. The small value of $\beta $ represents the high SI cancelation level. The system throughput of the non-cooperative scheme and the FCFS scheme have no change under different SI cancellation levels. The non-cooperative scheme has no V2V communications, so it does not need SI cancellation. In the FCFS scheme, vehicles mainly download content in the V2I phase, so the impact of SI cancellation in the V2V phase is not significant. With the improvement of SI cancelation level, the system throughput of the other two schemes all increase first and then keep stable. Properly increasing the SI cancelation level can improve transmission rates. When the SI cancelation level achieves a certain value (i.e., $\beta  = {10^{ - 13}}$ in Fig. \ref{fig11}), the system throughput of the proposed scheme and the random cooperation scheme start to be unchanged. In this case, the performance of these schemes mainly depends on different algorithms, and FD communications can no longer bring more promotion.

\section{Conclusion}\label{S7}

In this paper, we propose a content distribution scheduling scheme based on joint V2I and V2V communications in mmWave vehicular networks. In the V2I phase, the RSU selects vehicles that are conducive to content forwarding of the V2V phase to download the content according to the vehicular network topology and vehicle requirements. Further, in the V2V phase, FD communications and concurrent transmissions are exploited to improve transmission efficiency. Simulation results show that the proposed scheme reduces the number of transmission time slots for content distribution and significantly improves system throughput when compared with other schemes, especially under multiple vehicles and large-size files. In future work, we will consider the content distribution scheduling problem in the scenario with blocked mmWave links.

\bibliographystyle{IEEEtran}

\begin{thebibliography}{10}

\bibitem{5G}
M. Chen and Y. Wang, ``On the Key Technologies of Internet of Vehicles and Its Innovative Application of Integration with 5G,'' \emph{International Conference on Artificial Intelligence and Electromechanical Automation (AIEA)}, Tianjin, China, 2020, pp. 32-35.

\bibitem{Support}
J. Choi, V. Va, N. Gonzalez-Prelcic, R. Daniels, C. R. Bhat, and R. W. Heath, ``Millimeter-Wave Vehicular Communication to Support Massive Automotive Sensing,'' \emph{IEEE Communications Magazine}, vol. 54, no. 12, pp. 160-167, Dec. 2016.

\bibitem{Framework}
Z. Su, Y. Hui, and Q. Yang, ``The Next Generation Vehicular Networks: A Content-Centric Framework,'' \emph{IEEE Wireless Communications}, vol. 24, no. 1, pp. 60-66, Feb. 2017.

\bibitem{Challenges}
M. Giordani, A. Zanella, and M. Zorzi, ``Millimeter wave communication in vehicular networks: Challenges and opportunities,'' \emph{6th International Conference on Modern Circuits and Systems Technologies (MOCAST)}, Thessaloniki, Greece, 2017.

\bibitem{Challenges2}
K. Z. Ghafoor, L. Kong, S. Zeadally, A. S. Sadiq, and S. Mumtaz, ``Millimeter-Wave Communication for Internet of Vehicles: Status, Challenges, and Perspectives,'' \emph{IEEE Internet of Things Journal}, vol. 7, no. 9, pp. 8525-8546, Sep. 2020.

\bibitem{Networking}
K. Zheng, Q. Zheng, P. Chatzimisios, W. Xiang, and Y. Zhou, ``Heterogeneous Vehicular Networking: A Survey on Architecture, Challenges, and Solutions,'' \emph{IEEE Communications Surveys Tutorials}, vol. 17, no. 4, pp. 2377-2396, Fourth quarter 2015.

\bibitem{Data}
S. Ilarri, T. Delot, and R. Trillo-Lado, ``A Data Management Perspective on Vehicular Networks,'' \emph{IEEE Communications Surveys Tutorials}, vol. 17, no. 4, pp. 2420-2460, Fourth quarter 2015.

\bibitem{Taxis}
J. He, L. Cai, P. Cheng, and J. Pan, ``Delay Minimization for Data Dissemination in Large-Scale VANETs with Buses and Taxis,'' \emph{IEEE Transactions on Mobile Computing}, vol. 15, no. 8, pp. 1939-1950, Aug. 2016.

\bibitem{survey}
V. Va, T. Shimizu, G. Bansal, and R. W. Heath, Jr., ``Millimeter wave vehicular communications: A survey,'' \emph{Foundations and Trends in Networking}, vol. 10, no. 1, pp. 1-113, 2016.

\bibitem{CodeOn}
M. Li, Z. Yang, and W. Lou, ``CodeOn: Cooperative Popular Content Distribution for Vehicular Networks using Symbol Level Network Coding,'' \emph{IEEE Journal on Selected Areas in Communications}, vol. 29, no. 1, pp. 223-235, Jan. 2011.

\bibitem{Overview}
F. Arena and G. Pau, ``An Overview of Vehicular Communications,'' \emph{Future Internet}, vol. 25, no. 10, pp. 2362-2372, Oct. 2014.

\bibitem{D2D}
X. Cheng, L. Yang, and X. Shen, ``D2D for Intelligent Transportation Systems: A Feasibility Study,'' \emph{IEEE Transactions on Intelligent Transportation Systems}, vol. 16, no. 4, pp. 1784-1793, Aug. 2015.

\bibitem{Road}
A. Abdrabou and W. Zhuang, ``Probabilistic Delay Control and Road Side Unit Placement for Vehicular Ad Hoc Networks with Disrupted Connectivity,'' \emph{IEEE Journal on Selected Areas in Communications}, vol. 29, no. 1, pp. 129-139, Jan. 2011.

\bibitem{hoc}
T. Luan, X. Shen and F. Bai, ``Integrity-oriented content transmission in highway vehicular ad hoc networks,'' \emph{2013 Proceedings IEEE INFOCOM}, Turin, Italy, Apr. 2013.

\bibitem{WiFi-Based}
D. Zhang and C. K. Yeo, ``Enabling Efficient WiFi-Based Vehicular Content Distribution,'' \emph{IEEE Transactions on Parallel and Distributed Systems}, vol. 24, no. 3, pp. 479-492, Mar. 2013.

\bibitem{MMCD}
K. Ota, M. Dong, S. Chang, and H. Zhu, ``MMCD: Max-throughput and min-delay cooperative downloading for Drive-thru Internet systems,'' \emph{2014 IEEE International Conference on Communications (ICC)}, Sydney, NSW, Australia, 2014, pp. 83-87.

\bibitem{Multi-Hop}
J. Lv, X. He and T. Luo, ``Blockage Avoidance Based Sensor Data Dissemination in Multi-Hop MmWave Vehicular Networks,'' \emph{IEEE Transactions on Vehicular Technology}, vol. 70, no. 9, pp. 8898-8911, Sept. 2021.

\bibitem{Dynamic}
T. Wang, L. Song, Z. Han, and B. Jiao, ``Dynamic Popular Content Distribution in Vehicular Networks using Coalition Formation Games,'' \emph{IEEE Journal on Selected Areas in Communications}, vol. 31, no. 9, pp. 538-547, Sep. 2013.

\bibitem{Wang}
Y. Wang, H. Wu, Y. Niu, Z. Han, B. Ai, and Z. Zhong, ``Coalition Game Based Full-Duplex Popular Content Distribution in mmWave Vehicular Networks,'' \emph{IEEE Transactions on Vehicular Technology}, vol. 69, no. 11, pp. 13836-13848, Nov. 2020.

\bibitem{Chen}
L. Chen, Z. Li, and S. Jiang, ``Content downloading-oriented resource allocation joint scheduling in drive-thru networks,'' \emph{Ruan Jian Xue Bao/Journal of Software}, vol. 11, no. 2, pp. 27-38, First quarter 2014.

\bibitem{Graph}
K. Zheng, F. Liu, Q. Zheng, W. Xiang, and W. Wang, ``A Graph-Based Cooperative Scheduling Scheme for Vehicular Networks,'' \emph{IEEE Transactions on Vehicular Technology}, vol. 62, no. 4, pp. 1450-1458, May 2013.

\bibitem{Joint}
Q. Zheng, K. Zheng, P. Chatzimisios, and F. Liu, ``Joint optimization of link scheduling and resource allocation in cooperative vehicular networks,'' \emph{EURASIP J. Wireless Commun. Netw.}, vol. 2015, no. 1, pp. 170-183, Jun. 2015.

\bibitem{V2X1}
Y. Gao, X. Xu, Y. Guan and P. Chong, ``V2X Content Distribution Based on Batched Network Coding With Distributed Scheduling,'' \emph{IEEE Access}, vol. 6, pp. 59449-59461, Oct. 2018.

\bibitem{CVCG}
C. Chen, J. Hu, T. Qiu, M. Atiquzzaman and Z. Ren, ``CVCG: Cooperative V2V-Aided Transmission Scheme Based on Coalitional Game for Popular Content Distribution in Vehicular Ad-Hoc Networks,'' \emph{IEEE Transactions on Mobile Computing}, vol. 18, no. 12, pp. 2811-2828, Dec. 2019.

\bibitem{Offloading}
F. Malandrino, C. Casetti, C. Chiasserini, and M. Fiore, ``Offloading cellular networks through ITS content download,'' \emph{9th Annual IEEE Communications Society Conference on Sensor, Mesh and Ad Hoc Communications and Networks (SECON)}, Seoul, Korea (South), Jun. 2012, pp. 263-271.

\bibitem{Scenarios}
J Wang, K Liu, K Xiao, C Chen, W Wu, V. Lee, and S. Son, ``Dynamic Clustering and Cooperative Scheduling for Vehicle-to-Vehicle Communication in Bidirectional Road Scenarios,'' \emph{IEEE Transactions on Intelligent Transportation Systems}, vol. 19, no. 6, pp. 1913-1924, Jun. 2018.

\bibitem{Directional}
S. Pyun, W. Lee and D. Cho, ``Resource Allocation for Vehicle-to-Infrastructure Communication Using Directional Transmission,'' \emph{IEEE Transactions on Intelligent Transportation Systems}, vol. 17, no. 4, pp. 1183-1188, Apr. 2016.

\bibitem{Cooperative}
Q. Wang, P. Fan, and K. B. Letaief, ``On the Joint V2I and V2V Scheduling for Cooperative VANETs With Network Coding,'' \emph{IEEE Transactions on Vehicular Technology}, vol. 61, no. 1, pp. 62-73, Jan. 2012.

\bibitem{Throughput}
J. Chen, G. Mao, C. Li, A. Zafar and A. Y. Zomaya, ``Throughput of Infrastructure-Based Cooperative Vehicular Networks,'' \emph{IEEE Transactions on Intelligent Transportation Systems}, vol. 18, no. 11, pp. 2964-2979, Nov. 2017.

\bibitem{V2V/V2I}
Y. Ni, J. He, L. Cai and Y. Bo, ``Data Uploading in Hybrid V2V/V2I Vehicular Networks: Modeling and Cooperative Strategy,'' \emph{IEEE Transactions on Vehicular Technology}, vol. 67, no. 5, pp. 4602-4614, May 2018.

\bibitem{Full1}
Z. Xiao, P. Xia and X. Xia, ``Full-Duplex Millimeter-Wave Communication,'' \emph{IEEE Wireless Communications}, vol. 24, no. 6, pp. 136-143, Dec. 2017.

\bibitem{Full2}
G. Yang, M. Xiao, H. Al-Zubaidy, Y. Huang and J. Gross, ``Analysis of Millimeter-Wave Multi-Hop Networks With Full-Duplex Buffered Relays,'' \emph{IEEE/ACM Transactions on Networking}, vol. 26, no. 1, pp. 576-590, Feb. 2018.


\bibitem{UWB}
K. Liu, L. Cai and X. Shen, ``Exclusive-Region Based Scheduling Algorithms for UWB WPAN,'' \emph{IEEE Transactions on Wireless Communications}, vol. 7, no. 3, pp. 933-942, Mar. 2008.


\bibitem{STDMA}
J. Qiao, L. Cai, X. Shen and J. Mark, ``STDMA-based scheduling algorithm for concurrent transmissions in directional millimeter wave networks,'' \emph{2012 IEEE International Conference on Communications (ICC)}, Ottawa, ON, Canada, Jun. 2012.

\bibitem{Beamwidth}
J. Wildman, P. H. J. Nardelli, M. Latva-aho, and S. Weber, ``On the Joint Impact of Beamwidth and Orientation Error on Throughput in Directional Wireless Poisson Networks,'' \emph{IEEE Transactions on Wireless Communications}, vol. 13, no. 12, pp. 7072-7085, Dec. 2014.

\bibitem{Ding}
W. Ding, Y. Niu, H. Wu, Y. Li, and Z. Zhong, ``QoS-Aware Full-Duplex Concurrent Scheduling for Millimeter Wave Wireless Backhaul Networks,'' \emph{IEEE Access}, vol. 6, pp. 25313-25322, Apr. 2018.


\bibitem{MAC}
K. Tamaki, A. Raptino H., Y. Sugiyama, M. Bandai, S. Saruwatari and T. Watanabe, ``Full Duplex Media Access Control for Wireless Multi-Hop Networks,'' \emph{2013 IEEE 77th Vehicular Technology Conference (VTC Spring)}, Dresden, Germany, Jun. 2013.

\bibitem{RCTC}
W. Zhou, K. Srinivasan and P. Sinha, ``RCTC: Rapid concurrent transmission coordination in full DuplexWireless networks,'' \emph{2013 21st IEEE International Conference on Network Protocols (ICNP)}, Goettingen, Germany, Oct. 2013.

\bibitem{Nam}
C. Nam, C. Joo and S. Bahk, ``Joint Subcarrier Assignment and Power Allocation in Full-Duplex OFDMA Networks,'' \emph{IEEE Transactions on Wireless Communications}, vol. 14, no. 6, pp. 3108-3119, Jun. 2015.

\bibitem{Duplex1}
S. Goyal, P. Liu, O. Gurbuz, E. Erkip and S. Panwar, ``A distributed MAC protocol for full duplex radio,'' \emph{2013 Asilomar Conference on Signals, Systems and Computers}, Pacific Grove, CA, USA, Nov. 2013.

\bibitem{Niu1}
Y. Niu, L. Su, C. Gao, Y. Li, D. Jin, and Z. Han, ``Exploiting Device-to-Device Communications to Enhance Spatial Reuse for Popular Content Downloading in Directional mmWave Small Cells,'' \emph{IEEE Transactions on Vehicular Technology}, vol. 65, no. 7, pp. 5538-5550, Jul. 2016.

\bibitem{Relay}
Y. Wang, Y. Niu, H. Wu, B. Ai, Z. Zhong, D. O. Wu, and T. Juhana, ``Relay Assisted Concurrent Scheduling to Overcome Blockage in Full-Duplex Millimeter Wave Small Cells,'' \emph{IEEE Access}, vol. 7, pp. 105755-105767, Jul. 2019.


\end{thebibliography}

\end{document}